\documentclass[%
 aip,
 jmp,%
 amsmath,amssymb,
 reprint,%
]{revtex4-1}

\usepackage{graphicx}
\usepackage{dcolumn}
\usepackage{bm}

\usepackage{amsmath}
\usepackage{amssymb}
\usepackage{textcomp}
\usepackage{epstopdf}
\begin{document}


 \newcommand\psiinf{\psi_{\infty}}
 \newcommand\ub{\textbf{u}}
 \newcommand\Uhat{\widehat{U}}
 \newcommand\what{\widehat{\omega}}
 \newcommand\psihat{\widehat{\psi}}
 \newcommand\xhat{\widehat{x}}
 \newcommand\yhat{\widehat{y}}
 \newcommand\that{\widehat{t}}
 \newcommand\rhat{\widehat{r}}

\title{Start-up vortex flow past an accelerated flat plate}%

\author{Ling Xu}
 \email{lxu9@gsu.edu}
\affiliation{ 
Department of Mathematics and
Statistics, Georgia State University, Atlanta, GA 30303,USA
}%
\author{Monika Nitsche}%
 \email{nitsche@math.unm.edu}
\affiliation{ 
Department of Mathematics and Statistics, University of New Mexico,
Albuquerque, NM 87131, USA
}


\date{\today}

\begin{abstract}

Viscous flow past a finite flat plate moving in direction normal to itself
is studied numerically.
The plate moves with velocity $at^p$, where $p=0,0.5,1,2$.
We present the evolution of vorticity profiles, streaklines and streamlines,
and study the dependence on the acceleration parameter $p$.
%
Four stages in the vortex evolution, 
as proposed by Luchini \& Tognaccini (2002), 
are clearly identified. 
The initial stage, in which the vorticity consists solely of 
a Rayleigh boundary layer, is shown to 
last for a time-interval
whose length shrinks to zero like $p^3$, as $p \to 0$.
In the second stage, a center of rotation develops near the tip of 
the plate, well before a vorticity maximum within the vortex core develops.
Once the vorticity maximum develops, its position oscillates 
and differs from the center of rotation. 
The difference between the two 
increases with increasing $p$, and decreases in time.
In the third stage,
the center of rotation and the shed circulation closely satisfy 
self-similar scaling laws for inviscid flow. 
Finally, in the fourth stage, 
the finite plate length becomes relevant and
the flow begins to depart from the self-similar behaviour. 
While the core trajectory and circulation closely satisfy inviscid scaling laws,
the vorticity maximum and the boundary layer thickness follow
viscous scaling laws.
The results are compared with experimental results of Pullin \& Perry (1980),
and Taneda \& Honji (1971), where available.
%
%
%

\end{abstract}


\maketitle

\section{Introduction}

This paper presents numerical simulations of the starting vortex flow at the edge of 
an accelerating finite flat plate. The plate
is assumed to have zero thickness, 
and moves with speed $U(t)=at^p$ in direction normal to itself.
The flow is nondimensionalized based on the plate length and the parameter $a$,
yielding a characteristic flow Reynolds number $Re$.
Here we 
study the effect of the acceleration parameter $p$ for 
fixed Reynolds number $Re=500$. The effect of varying $Re$ 
for fixed $p$ is presented elsewhere \cite{xunitsche14}. 

Being of intrinsic interest in fluid dynamics,
the starting vortex flow 
has been the focus of many research works,
beginning with the work of Prandtl in 1904 \cite{lugt96,anderson05}.
Most relevant to the accelerated case considered here are the following.
Taneda \& Honji \cite{taneda71} presented experimental results for
both uniform and 
accelerated flow past a plate, and observed an apparent scaling
of the vortex size. 
Pullin and Perry \cite{pullinperry80} performed experiments 
of flow past finite wedges, with wedges as small as $5^{\circ}$,
which serve as a basis of comparison for our present results. 
They reported detailed measurements of the vortex core trajectory
at early times, and compared their  observations to inviscid similarity theory
results obtained by Pullin \cite{pullin78}.
The theory holds for inviscid vortex sheet separation at the
edge of a semi-infinite plate.
In view of the absence of a plate or viscous length scale,
in that case the vortex center trajectory and the shed 
circulation satisfy inviscid scaling laws in time that were
already reported by Kaden\cite{kaden31} in 1931. 
Pullin\cite{pullin78} computed the time-independent 
self-similar shape using an iterative scheme. 

Scaling laws also exist for viscous flow past a semi-infinite
plate, and follow from dimensional analysis. 
However, in this case the solution depends on the viscous length
scale. The viscous scaling was exploited by 
Luchini and Tognaccini\cite{luchini02}, 
who computed flow past a semi-infinite plate
in a self-similar reference frame. 
The finite plate case was studied numerically by 
Koumoutsakos and Shiels \cite{koushiels96}. 
They computed flow past a plate moving with either 
impulsively started velocity or constant acceleration.
For their accelerated flow, they observe a shear layer instability similar to 
that observed by Pierce \cite{pierce61} and Lian and Huang \cite{lianhuang89}.
These numerical results are given mostly for relatively large times.

Here, we use highly resolved simulations to present a systematic
study of the dependence on $p$ over a large range of times.
We present the evolution of vorticity profiles,
streaklines and streamlines, track vortex core trajectories and vorticities,
and compute the shed circulation following the approach taken in Xu and Nitsche\cite{xunitsche14}.

Luchini and Tognaccini\cite{luchini02} propose 
four different time regimes for viscous flow past finite plates,
and it is interesting to identify these regimes using the present simulations.
We observe and present the timescales of an initial Rayleigh flow regime.
We study the emergence of a vortex core in a second regime, and 
compare two possible definitions, namely the center of rotation 
and the position of the vorticity maximum.
We observe the self-similar scaling laws in a third regime, 
and estimate the time at which the finite plate length dominates,
in a fourth regime.
All results are computed with $Re=500$, with the exception of 
a comparison with experimental results by Pullin and Perry,
for which $Re=6000$, and $p=0.45$.
The results are also compared with the scaling behaviour proposed by 
Taneda and Honji\cite{taneda71}.

The paper is organized as follows. Sections II and III present the problem and
the numerical method used, section IV presents the numerical results,
section V summarizes the observations.

\section{Problem formulation}\label{sec:problem}

\begin{figure}
\centering
\includegraphics[width=0.45\textwidth]{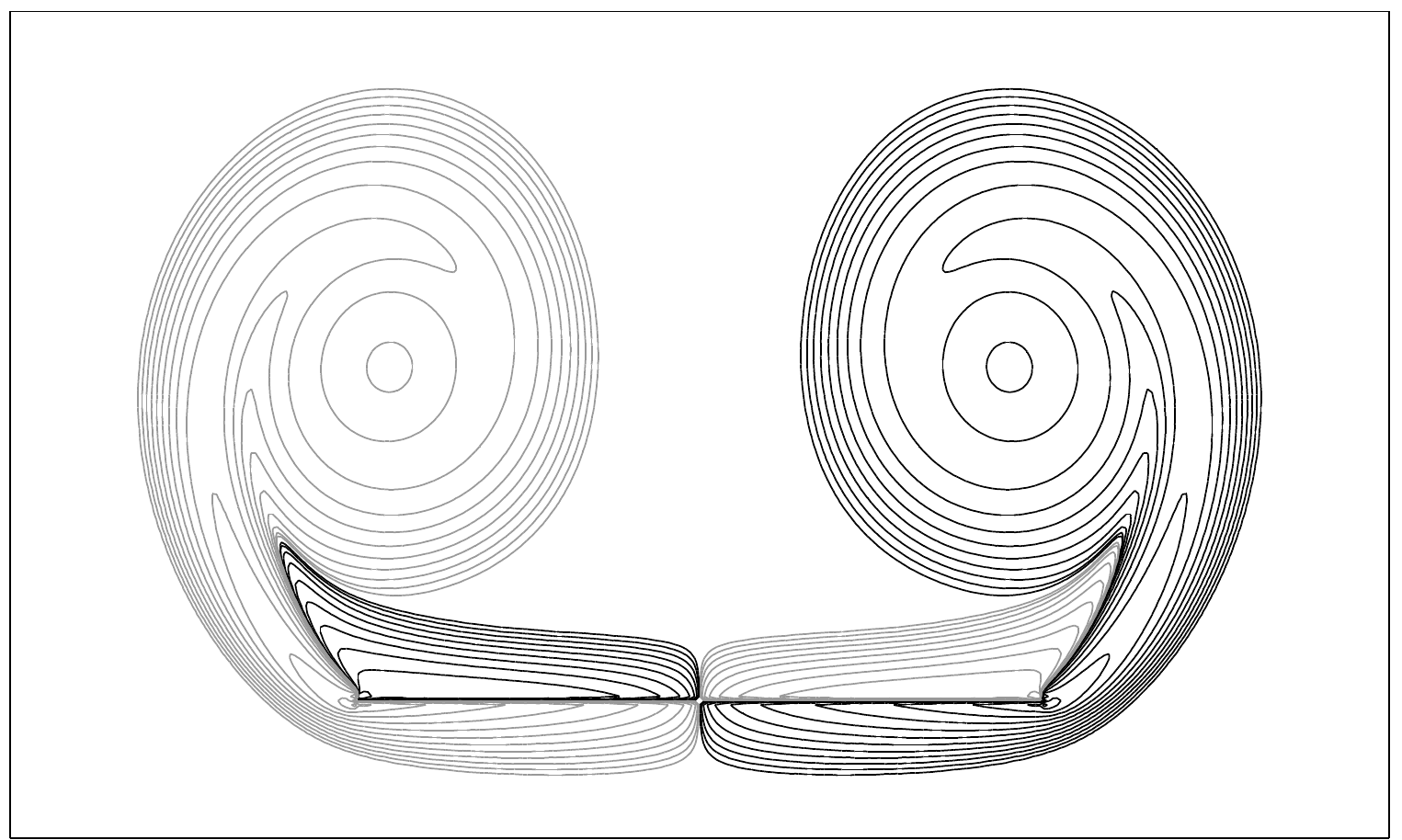}
\caption{Sample vorticity contours for $p=0$ at a
relatively large time, with positive vorticity in black, negative vorticity in grey.}
\label{fig1}
\end{figure}

A finite plate of length $L$ and zero thickness, immersed 
in a homogeneous viscous fluid, is accelerated in direction normal 
to itself with speed 
\begin{equation}\label{eqn:farfield}
\Uhat(\that)=a \that^p~,
\end{equation}
where $a$ is a dimensional constant.
Here and throughout the paper, the hat symbol denotes that the 
variables are dimensional.
We consider $p$ = 0, 0.5, 1, 2. 
These include impulsively started flow ($p=0$),
uniform acceleration ($p=1$), and linear acceleration ($p=2$).
The flow is assumed to be two-dimensional, 
and to remain symmetric about the centerline 
at all times. 
We choose a reference frame fixed on the plate, in which the plate is horizontal
and the driving velocity moves upwards, approaching 
parallel flow in the far field. 
Figure 1 illustrates the fluid vorticity some time after the beginning of
the motion. As fluid moves from upstream (below the plate) to downstream
(above the plate), boundary layers of vorticity form along the plate
walls which eventually separate and form a pair of 
counterrotating vortices.
We note that changing to an accelerated time-dependent reference frame 
does not affect the vorticity dynamics, but only 
the pressure and thus the forces acting on the plate. 
The results below are therefore the same, up to a translation,
as those for a moving plate in a reference frame fixed at infinity. 

The flow is nondimensionalized with respect to the plate length
and the parameter $a$, yielding a characteristic timescale
 \begin{eqnarray}
  T = \left(\frac{L}{a}\right)^{{1}/({p+1})}~,
 \end{eqnarray}
and flow Reynolds number 
\begin{eqnarray}\label{re}
Re = \frac{L^2}{\nu T}
=\frac{a^{1/(p+1)}L^{(2p+1)/(p+1)}}{\nu}~,
\end{eqnarray}
where $\nu$ is the kinematic fluid viscosity.
The problem is described in nondimensional time $t$ and 
Cartesian coordinates $(x,y)$, 
chosen so that the plate lies on the x-axis, centered at the origin,
at $\{(x,0)\,|\,x\in[-1/2,1/2]\}$. 
The fluid velocity and scalar vorticity are $\left(
u(x,y,t),v(x,y,t)\right) $ and $\omega(x,y,t)$. 
The relation between dimensional and nondimensional variables is,
for example
\begin{equation}
t={\that\over T}\,,~
x={\xhat\over L}\,,~
U={\Uhat T\over L}\,,~
\omega={\what T}\,.
\end{equation}
The nondimensional far field velocity is $(0,t^p)$.

The flow is governed by the incompressible Navier-Stokes equations,
\begin{eqnarray}\label{nse}
\frac{D \omega}{D t} &=& {1\over Re}\nabla^{2}\omega~,\nonumber\\
\nabla^{2} \psi &=& -\omega~,\\
\left(u,~v\right) &=& \nabla^{\perp}\psi = \left(\frac{\partial\psi}{\partial y},
-\frac{\partial\psi}{\partial x} \right)~.\nonumber 
\end{eqnarray}
It is initially irrotational,
$\omega(x,y,0)=0$,
with boundary conditions 
$\psi=0$ and $u=0$ on the plate,
 and $\psi(x,y,t) \to \psiinf$ as $|(x,y)|\to\infty$.
Here, $\psiinf$ is the potential flow that induces
the far field velocity, 
given by the complex potential 
\begin{equation}
W_{\infty}(x,y,t)=t^p\sqrt{{1\over4}-z^2}=\phi_{\infty}+i\psiinf~,
\end{equation}
where $z=x+iy$. 

\begin{table*}[ht!]
\begin{ruledtabular}
\caption{\label{comp_p} List of parameters in computations: the mesh
size $h$, time step $\Delta t$, 
starting/ending time $t_{\text{start}}$/$t_{\text{end}}$, 
computational domain $[0, x_{\text{max}}$]$\times$[$y_{\text{min}},
y_{\text{max}}]$, the number of grid
points $N_x\times N_y$, the value(s) of $p$ and the Reynolds number $Re$.}
 \begin{tabular}{lclllll}
  $h$ & $\Delta t$ & [$t_{\text{start}}$, $t_{\text{end}}$] &
$[0, x_{\text{max}}$]$\times$[$y_{\text{min}}, y_{\text{max}}$] &
$N_x\times N_y$ & $p$ & $Re$\\ 
  1/5120 & $2\times 10^{-6}$ & [0, 0.0004] & 
[0, 0.55]$\times$[-0.05, 0.1] & 2816$\times$768 & 0 &$500$\\ 
  1/2560 & $4\times 10^{-6}$  & [0, 0.005]  & 
[0, 0.55]$\times$[-0.05, 0.1] & 1408$\times$384 & 0&$500$\\ 
  1/1280 & $5\times 10^{-5}$      & [0, 0.1]&
[0, 0.75]$\times$[-0.125, 0.25] & 960$\times$480 & 0, 0.5, 1, 2&$500$\\ 
1/640 & $1\times10^{-4}$      & [0.1, 0.7]&
[0, 0.1]$\times$[-0.25, 0.75] & 480$\times$480 &0, 0.5, 1, 2&$500$\\ 
  1/320 & $2\times 10^{-4}$     & [0.7, 3]&
[0, 1.5]$\times$[-0.25, 2.75] & 480$\times$960 &0, 0.5, 1, 2&$500$\\  
  1/160 & $4\times 10^{-4}$     & [3, 4]&
[0, 1.5]$\times$[-0.5, 5.5] & 240$\times$960 &0 &$500$\\ 
\multicolumn{6}{c}{} \\
  1/1280 & $5\times 10^{-5}$      & [0, 0.75]&
[-0.1, 0.4]$\times$[0, 0.5] & 960$\times$640 & 0.45&$6000$\\ 
 \end{tabular}
\end{ruledtabular}
\end{table*}


 \section{Numerical Method} 

The governing equations (\ref{nse}) are solved in 
a finite rectangular domain in
the right half plane, $[0,x_{max}]\times[y_{min},y_{max}]$, using a time-splitting 
mixed finite-difference and semi-Lagrangian scheme. 
The domain in space and time is discretized using a uniform mesh, 
$\Delta x=\Delta y=h$, with constant timestep $\Delta t$,
over a given time interval.
The solution is advanced from time $t_n$ to $t_{n+1}$ by convecting the
current vorticity according to 
\begin{equation}
{D\omega\over Dt}=0
\end{equation}
using a semi-Lagrangian scheme; using the updated vorticity to obtain 
updated interior and  boundary streamfunction, velocity and vorticity
values; and then solving 
\begin{equation}
{\partial\omega\over \partial t}={1\over Re} \nabla^2\omega
\end{equation}
using an implicit Crank-Nicolson method.
The method is described in detail in Xu \& Nitsche\cite{xunitsche14}
and is based on the work in Xu\cite{xu12}.

Table \ref{comp_p}
lists all parameters used in the present computations for various values of $p$. 
In several cases, 
in order to compute the flow to large times, 
the computations were performed using a fine
mesh until some early time, subsampling that result
and continuiung on a coarser mesh, and repeating this process. 
The time intervals used are indicated in the table. For example, 
the result for $p=0$ at $t=4$ is obtained using $h=1/1280$ for $t\in[0,0.1]$,
subsampling the result at $t=0.1$ to continue with $h=1/640$ until $t=0.7$, etc. 
%
The case of $p=0$ is more difficult
to compute in view of the initial singularity at the plate tip 
discussed in \cite{xunitsche14}. Therefore, finer resolutions are used in this
case than for $p>0$.


In order to compare with laboratory experiment, we also compute
particle streaklines by releasing a fluid particle into the flow from
the plate tip at each timestep, and evolving it with the fluid velocity.

\section{Numerical results} \label{sec:results}

\subsection{Vorticity, streamlines, and streaklines for $p=1$}\label{p1}
  \begin{figure*}
  \centering
  \includegraphics[width=0.75\textwidth]{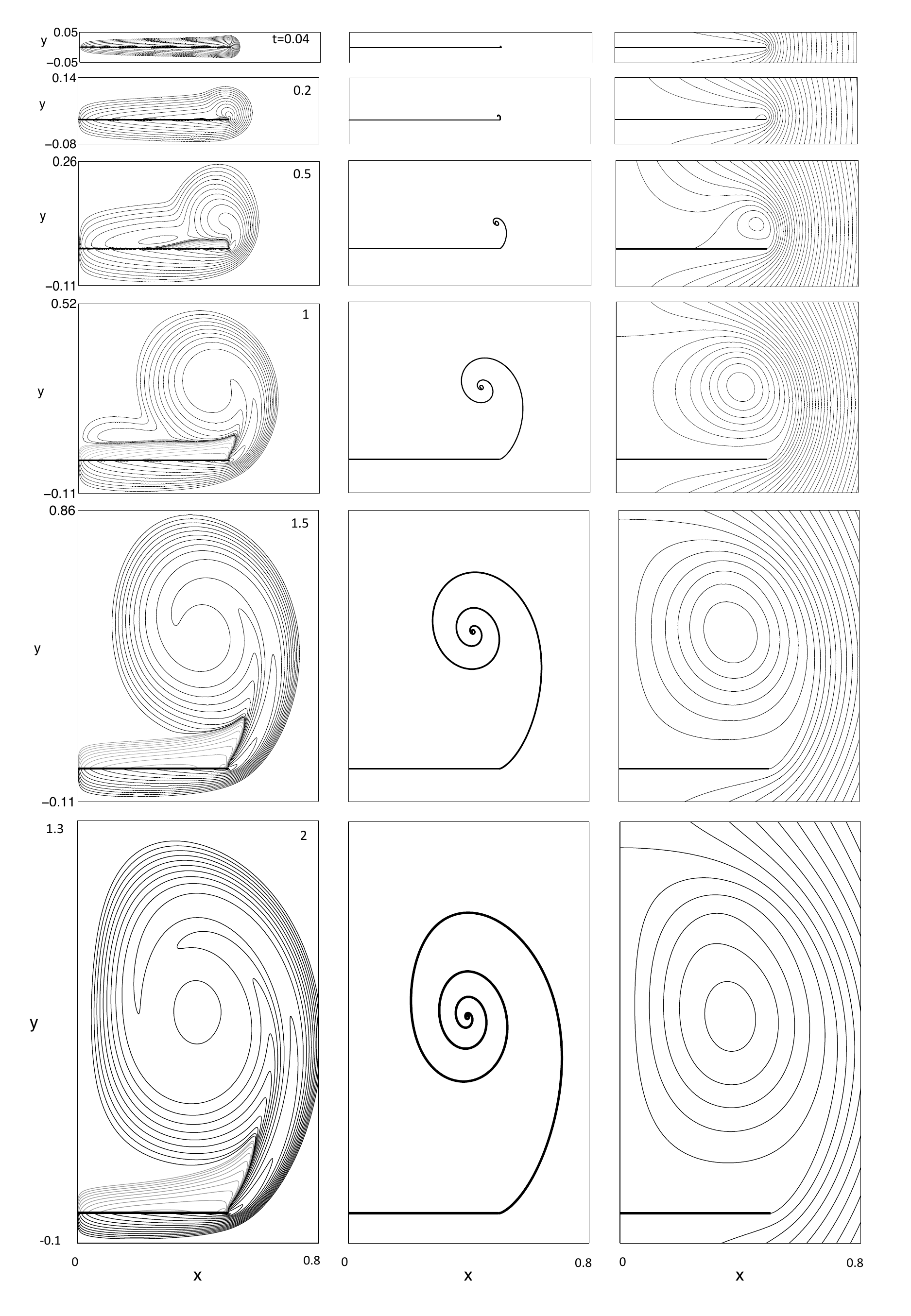}
  \caption{
  Vorticity (left), streaklines (middle) and streamlines (right) at
  $t$= 0.04, 0.2, 0.5, 1, 1.5, 2, for $p=1$, $Re=500$.
   Vorticity contours level are $\pm 2^j$, $j=-5,\dots,12$.}
  \label{F:evol}
  \end{figure*}

  \begin{figure*}
  \includegraphics[width=0.8\textwidth]{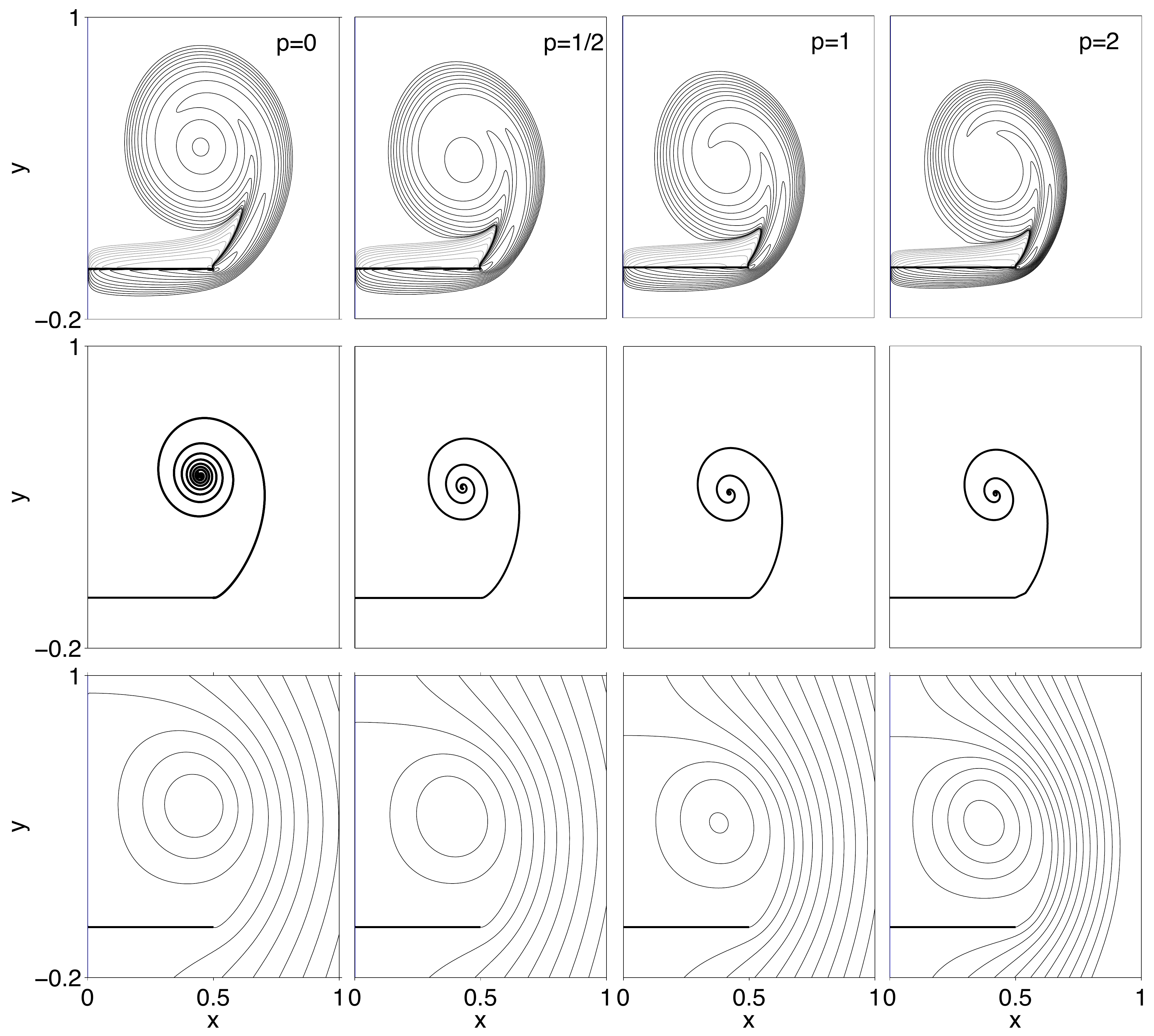}
  \caption{Vorticity(\textit{top}),
  streaklines(\textit{middle}) and
  streamlines(\textit{bottom}) for fixed 
  displacement $d=1$,   and $p=0,1/2,1,2$, as indicated.}
  \label{F:allp}
  \end{figure*}

  \begin{center}
  \begin{figure}
  \includegraphics[width=0.35\textwidth]{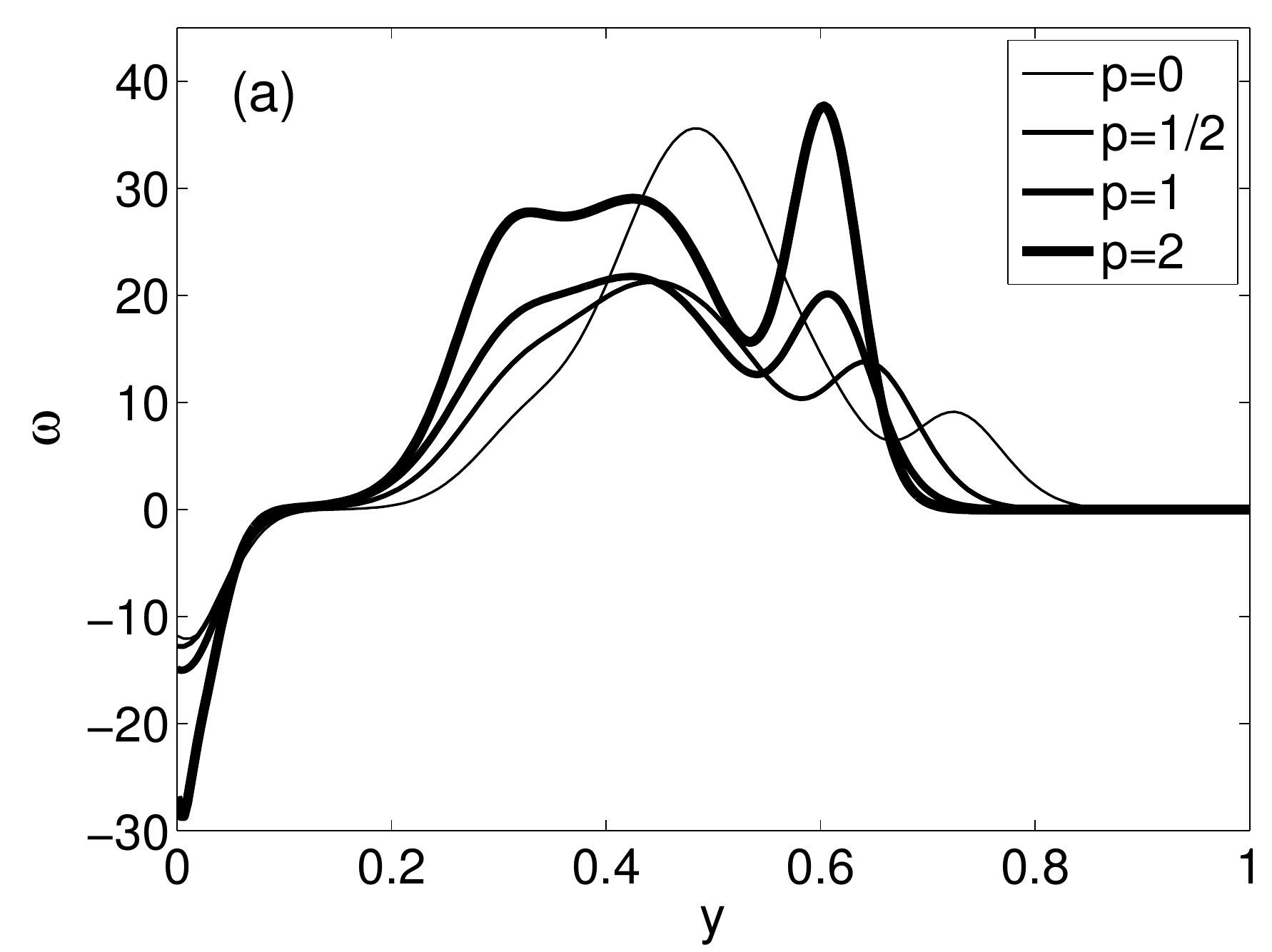}
\hbox{ }\hskip2pt
  \includegraphics[width=0.34\textwidth]{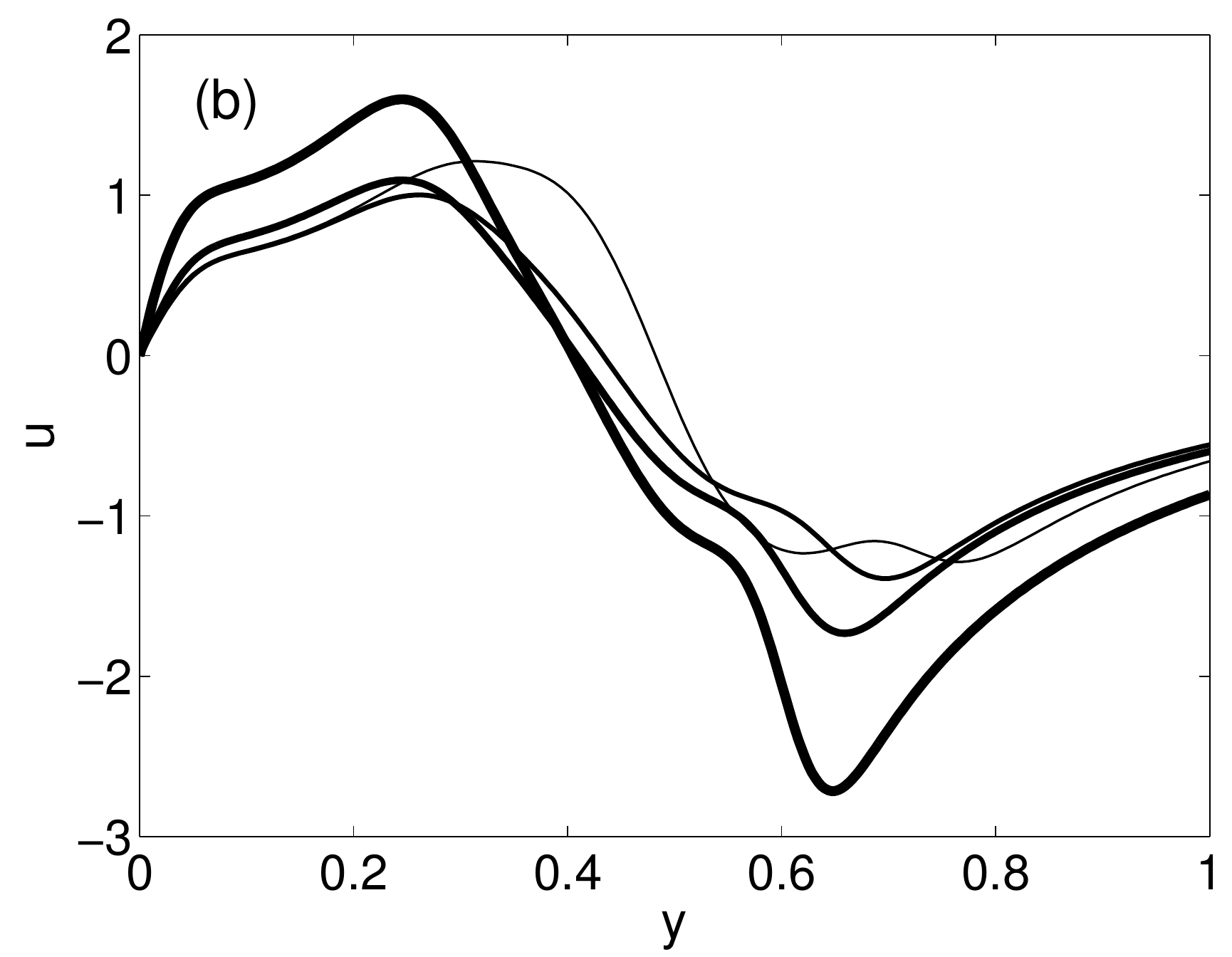}
  \caption{(a) Vorticity and (b) velocity at $d=1$ along a vertical line $x=x_m$ 
through the vorticity maximum, vs. $y$, for $p=0,1/2,1,2$, as indicated. 
}
  \label{F:profile}
  \end{figure}
  \end{center}

  \begin{figure}
  \includegraphics[width=0.5\textwidth]{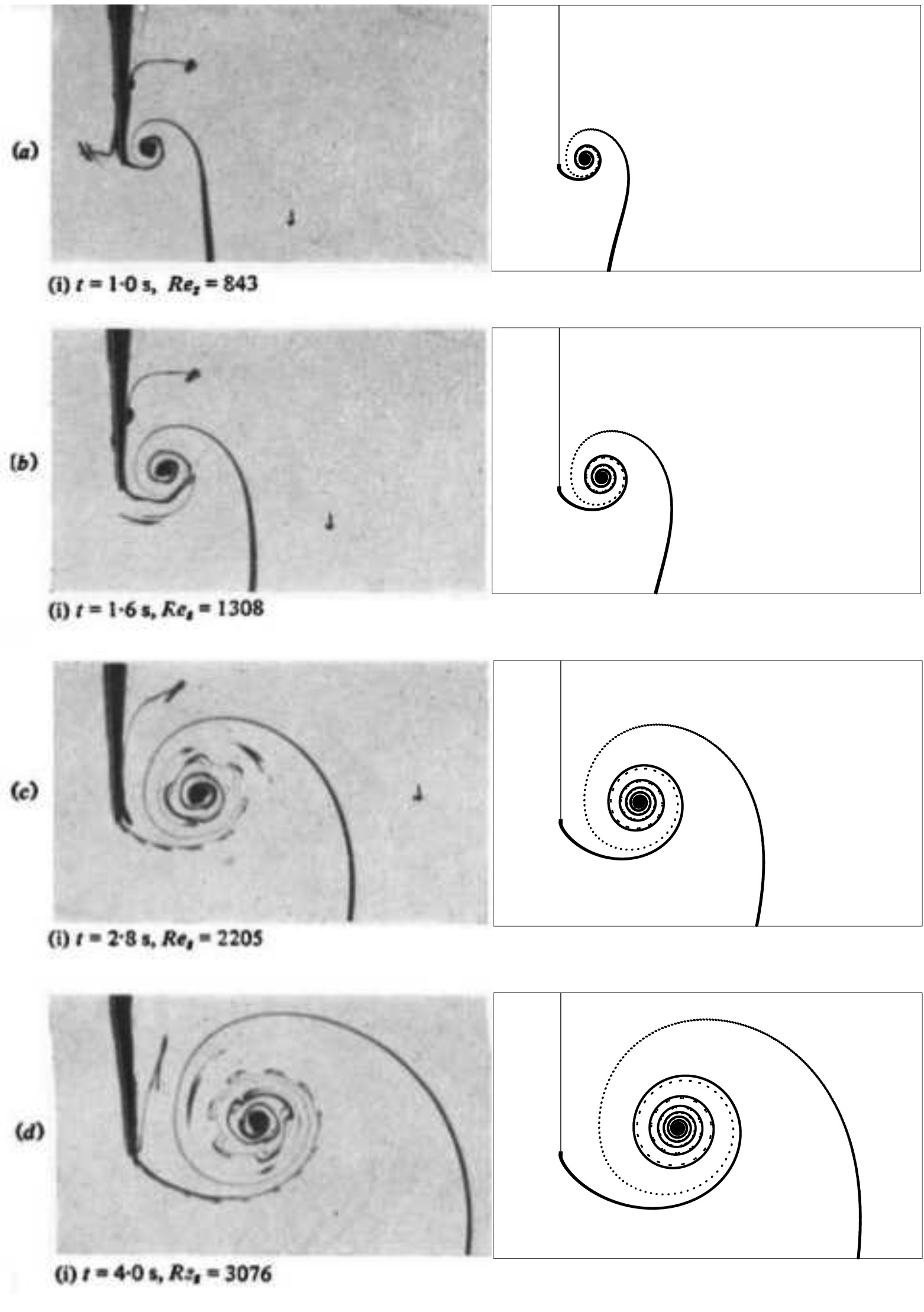}
  \caption{Left column: Streaklines for flow past a wedge of 
angle $\beta=5^o$, at $t$=1s, 1.6s, 2.8s and 4s, for $p=0.45$ and $Re=6621$, 
obtained by Pullin and Perry\cite{pullinperry80} from laboratory experiments 
(reproduced with permission from the Journal of Fluid Mechanics).
Right column: Numerical simulations for flow past a plate ($\beta=0^o$)
at the same times, for $p=0.45$ and $Re=6000$.}
  \label{F:pullin_streak}
  \end{figure}

Figure \ref{F:evol} presents the computed flow evolution 
for $p=1$ (constant acceleration) and $Re=500$.   
The figure shows vorticity contours (left), streaklines (middle), and streamlines (right) at
$t=$0.04, 0.2, 0.5, 1, 1.5, 2, as indicated.
Positive vorticity contours are shown in black, negative ones in a
lighter shade of grey.
Results are plotted in the right half plane, $x\ge 0$. By symmetry, the
vorticity and streamfunction in $x\le0$ are of equal magnitude but opposite sign.

Initially, the background flow has zero velocity. As the motion begins,
boundary layers of positive vorticity form around the plate,
on both the upstream side (below)
and downstream side (above).
This early vorticity is almost symmetric across the plate at $y=0$,
as can be seen in figure \ref{F:evol} at $t=0.04$. 
The corresponding streamlines, in the right column, are also practically
symmetric across the plate. It is noteworthy to remark that
this almost symmetric flow regime is not present for the impulsively
started case presented in \cite{xunitsche14}. This will be discussed in more
detail below.

As time evolves, vorticity upstream from the
plate is convected downstream, concentrating near the tip
as a vortex that grows in time and breaks the approximate symmetry of 
the initial boundary layers.
At the same time that the vorticity concentrates, 
a region of recirculating flow forms near the tip. 
The recirculating region is bounded by the zero level
streamline and is seen in figure \ref{F:evol} 
as early as $t=0.2$, in the right column.
Within the recirculation region and close to the wall, fluid 
moves in the opposite
direction of the background flow, thereby 
generating a region of opposite signed
boundary layer vorticity.
This negative wall vorticity appears in the fluid flow at $t\approx$0.1, 
and can be seen in figure \ref{F:evol} for all times $t\ge 0.2$.
The appearance of the recirculation region near the tip 
and the associated concentration of vorticity marks
the formation of the starting vortex.  

The upstream boundary layer vorticity 
keeps being convected downstream, feeding the starting
vortex. As a result, the leading vortex grows in size, 
and induces more negative
vorticity on the plate.
The negative vorticity
layer thickens and extends toward the axis at $x=0$, together
with the recirculation region.  
At $t$=1, the downstream wall vorticity is all negative, 
and separates the plate from the original positive 
boundary 
vorticity.
The remaining positive boundary vorticity diffuses and vanishes, 
as seen here at times $t=1.5, 2$, 
leaving the starting vortex clearly separated from the original
boundary layer vorticity.
The negative vorticity near the tip is entrained by the vortex,
which grows and convects in the downstream direction.

%

The streaklines shown in the middle column in figure \ref{F:evol}
are computed by releasing fluid particles from the plate tip at each time step,
and computing their evolution with the fluid velocity. 
At any given time, the figure \ref{F:evol} shows the current position of all 
particles released previously, mimicking what can be visualized in
laboratory experiments. The particles rotate around the vortex center,
with particles that have been released earlier traveling closer to the center
than those released later. This gives the 
resulting streaklines their spiral shape.   
We note that the tip, which is the point at which the particles are released, 
is also the point at which the flow vorticity is maximal. 
The particles thus approximate the convection of the maximum vorticity,
and thus approximate the centerline of maximum vorticity in
the separated shear layer. 
However, the size of the spiral streakline is not representative of 
the size of the vortex recirculation region.

\subsection{Dependence on $p$ at fixed displacement $d=1$}\label{d1}
The parameter $p$ describes the driving
velocity $(0,t^p)$ in the far field.
In a reference frame fixed at infinity, the plate moves 
downward with velocity $(0,-t^p)$.
The solution for varying $p$ at a fixed time $t$
varies greatly since the plate has travelled 
significantly different distances $d={1\over p+1}t^{p+1}$ 
at equal time $t$, resulting in 
vortices of significantly different size.
It is more meaningful 
to compare solutions with varying $p$ 
at times at which the plate displacement $d$ is equal.
All results shown herein that compare the solution for various 
$p$ are therefore 
plotted in reference to the displacement $d$, instead of time $t$.

Figure \ref{F:allp} compares the vorticity profiles
(top), streaklines (middle) and flow streamlines (bottom) at fixed
displacement $d=1$ for all $p$=0, 1/2, 1, 2 computed, as indicated. 
As $p$ increases, the vorticity contours show that the
vortex decreases slightly in size, and the core vorticity becomes
more uniform. 
Furthermore, the wall boundary layer thickness decreases slightly.
The vorticity profile as a function of $p$ is more clearly shown 
in figure \ref{F:profile}(a),
which plots the vorticity at $d=1$ along the vertical line $x=x_m$ through
the vorticity maximum $(x_m,y_m)$ in the vortex center, as a function of $y$,
for all $p$ computed. It shows that the profile is flatter near the core
for larger values of $p$. The outermost shear layer turn is stronger for larger
$p$, with larger maximal vorticities, 
and peaks at a smaller value of $y$, reflecting the smaller shape
seen in the vorticity contours.

The spiral streaklines plotted in the middle row of figure \ref{F:allp}
show that as $p$ increases, the spiral size decreases,
and the roll-up away from the center is less tight.
That is, for larger $p$, more particles released at early times
end up near the spiral center. 
This is caused by the fact that 
for larger $p$, the vortex has travelled
less far from the plate at early times.
As $p$ increases, the spiral shape is more elliptical and less round. 
It also leans further to the left, 
with the line from the spiral center to the plate tip 
subtending a smaller angle with the plate. 

The streamlines plotted in the bottom row in figure \ref{F:allp}
show that as $p$ increases, the size 
of the recirculation region decreases.
From the streamline density we deduce that the velocity gradients
in the core first
decreases as $p$ increases from $0$ to $p=1/2$, but then 
increase as $p$ increases past 1/2.
Figure \ref{F:profile}(b) plots the velocity along $x=x_m$,
and shows that this is indeed the case. The profiles also show
the decreasing value of $y$ of the point with $u=0$, near the rotation center.

\subsection{Comparison with laboratory experiments}\label{p045}
The most detailed experimental results available for comparison 
are those of Pullin \& Perry\cite{pullinperry80}
of flow past a wedge of angle $\beta$.
Their streakline visualization for their smallest wedge angle used,
$\beta=5^o$, is shown in figure \ref{F:pullin_streak}, left column.
The photographs show snapshots of an experiment performed in a 
rectangular tank, in which water flows from 
left to right past a planar wedge 
of height $h=12.7$cm
attached to the top of the tank. Near the midsection of the tank,
the flow is close to planar. 
Dye initially lines the walls of the wedge near its tip, and trickles almost
vertically downward before the motion begins.
The photographs show the position of the dye at the indicated dimensional
times after the begin of the fluid flow, for (dimensional) 
background velocity $U=at^p$, with $p=0.45$, 
$a=0.86$ cm/sec$^{p+1}$. 
By symmetry, the flow is comparable to flow past a
plate with $L=2h$. At water temperature of $24^o$C, 
the corresponding Reynolds number
as defined in equation (3) above is $Re=6621$.
 
The experimental results are compared with
the present numerical simulations 
of flow past a plate ($\beta=0^o$), with $p=0.45$ and $Re=6000$.
The right column in figure \ref{F:pullin_streak} shows the 
computed streaklines
at the nondimensional times corresponding to those shown in the left column,
using the timescale in equation (2).
The computed results are shown 
in a rotated frame, for better comparison with the experiments.
The figure also plots the position of particles initially
placed along a vertical line below the plate (in the rotated frame)
to better reproduce the experiment.
These particles form the outermost turn of the spiral streakline.
  \begin{figure*}
  \centering
  \includegraphics[width=1.01\textwidth]{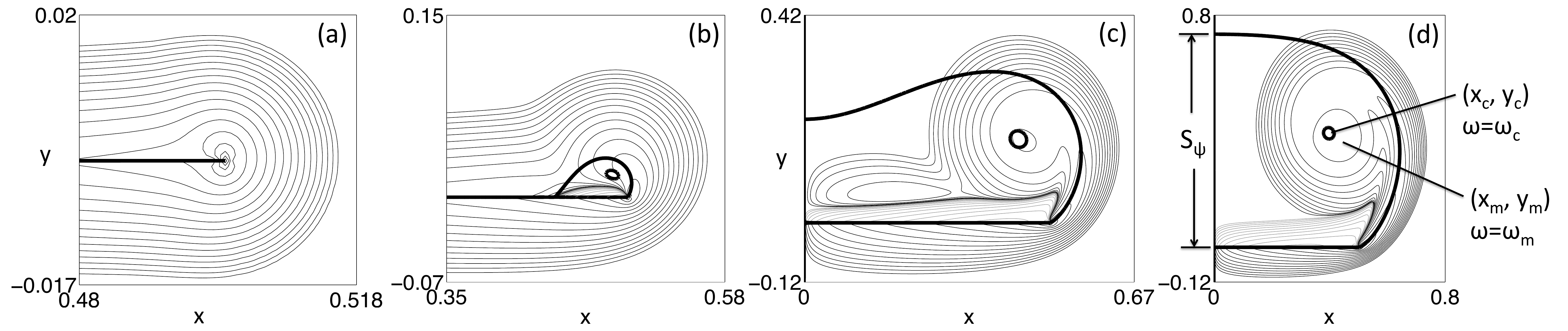}
  \caption{Four stages in the vorticity evolution, illustrated here for $p=1/2$,
at times (a) $d=0.00024$ ($t=0.005$), (b) $d=0.024$ ($t=0.11$), 
(c) $d=0.310$ ($t=0.6$) and (d) $d=0.88$ ($t=1.2$). 
Vorticity contours and two streamlines (thick solid lines) are shown.
The center of rotation $(x_c,y_c)$, the position of the vorticity 
maximum $(x_m,y_m)$,
the corresponding vorticities $\omega_c$ and $\omega_m$, and the
vortex size $s_{\psi}$ are indicated in plot (d).}
  \label{F:stages}
  \end{figure*}

\begin{center}
  \begin{figure}
  \includegraphics[width=0.4\textwidth]{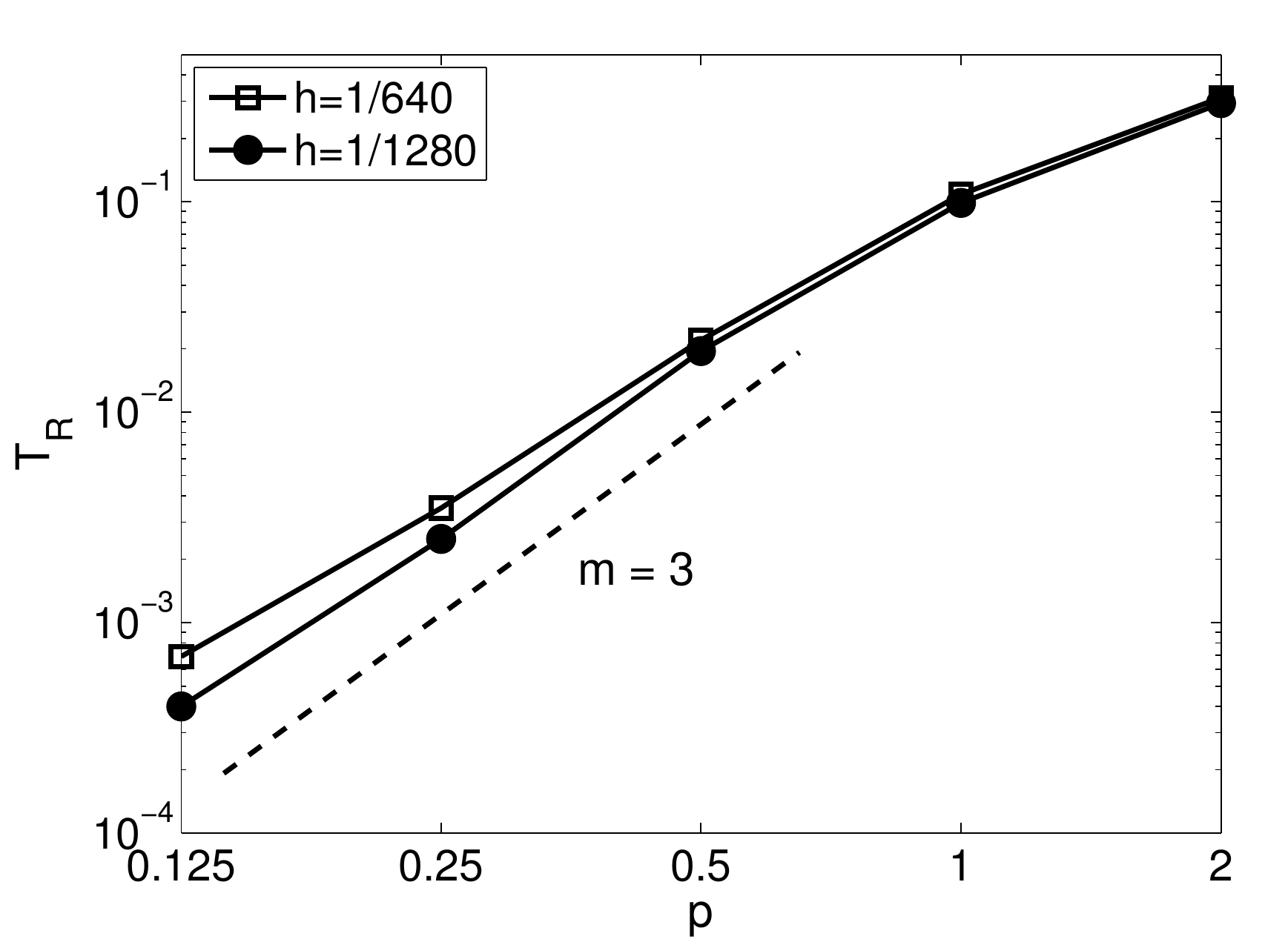}
  \caption{Duration $T_R$ of the Rayleigh stage, as a function of $p$,
computed with two resolutions $h=1/1280$ and $h=1/640$. 
The dashed line has the indicated slope $m$.}
  \label{F:tr}
  \end{figure}
\end{center}

Good agreement between experimental and numerical results is observed for 
the spiral streakline size, the overall spiral shape, and the 
spiral center position.
The spiral centers will be compared in more detail
in section E below.
One difference is observed in the spiral turn emanating from
the plate tip, which displays small vortices in the experiment, 
reflecting an instability that is not seen in the computed results 
at these times.
Schneider \textit{et al}\cite{schneider14} give evidence that 
the shape of a finite thickness plate tip can contribute to these 
oscillations.  
Another difference is observed in the outer spiral turn, which
has moved further to the right in the experiments than in the computations.
This difference may be due to differences in the initial particle positions
below the plate, for which the experimental data is not available.
It may also be due to differences in the wedge angle between
the experiment and the computation, whose effects remains to be studied.

\subsection{Four stages in the vortex evolution}\label{4stages}
Luchini and Tognaccini\cite{luchini02} propose four stages in the
evolution of the starting vortex. 
It is interesting to identify them here, as illustrated in figure
\ref{F:stages}. The
figure plots vorticity contours for $p=1/2$ at an increasing
sequence of displacements (see caption). 
It also plots the level curve $\psi=0$ 
that bounds the region of recirculating flow, as 
well as a streamline close to the center of rotation within this
region.
  \begin{figure}
  \centering
  \includegraphics[width=0.47\textwidth]{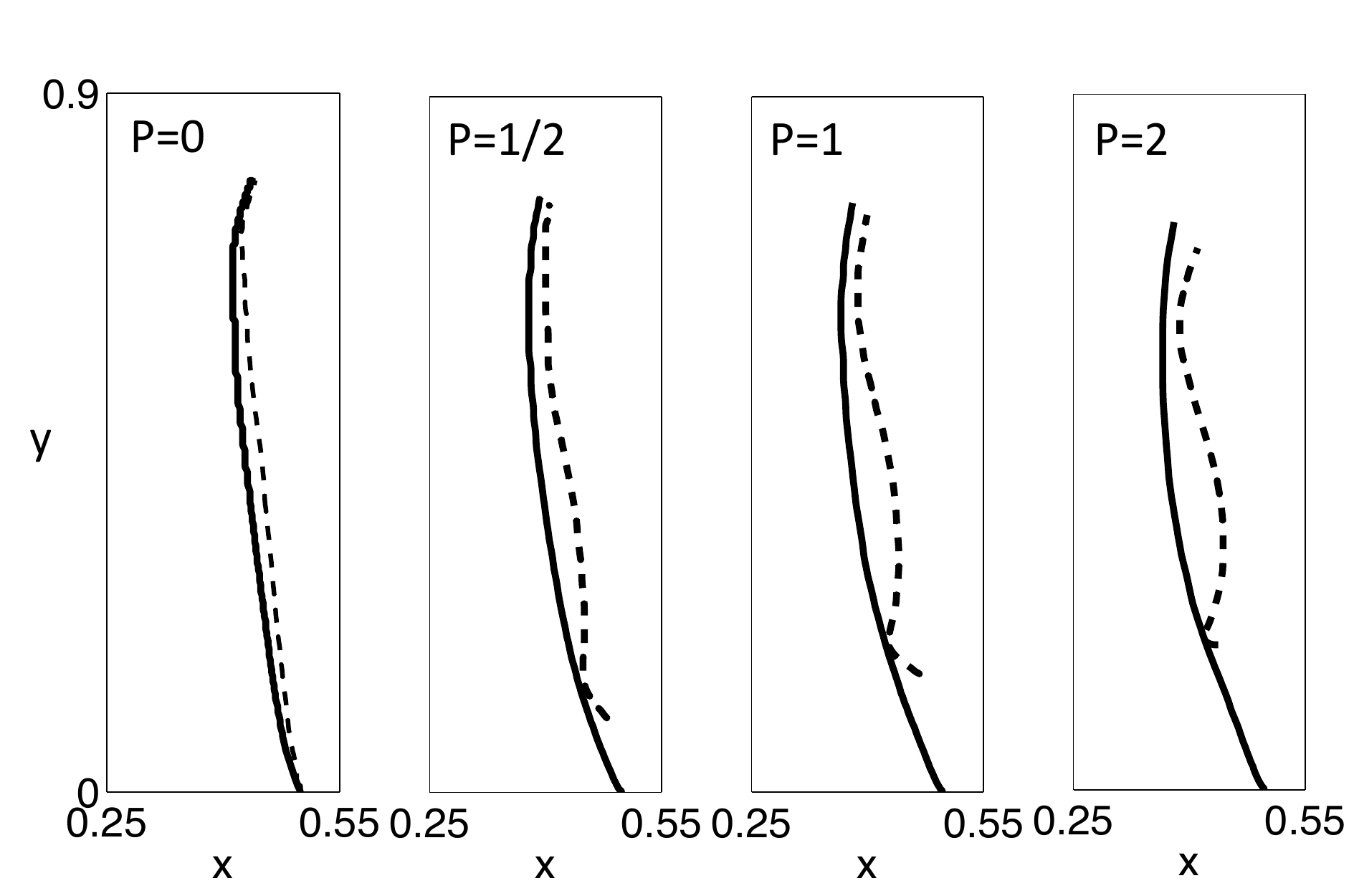}
  \caption{Trajectories of the rotation center $(x_c,y_c)$ (solid line)
and the position of the vorticity maximum $(x_m,y_m)$ (dashed line) 
for various $p$, as indicated, for $d\in[0, 3]$.}
  \label{F:comptraj}
  \end{figure}

In the first stage, referred to as the Rayleigh stage and 
illustrated in figure \ref{F:stages}(a), 
the vorticity consists of an almost symmetric boundary 
layer of uniform thickness around the whole plate, without any
apparent separated flow. 
In the second stage a region of recirculating flow
has formed near the tip of the plate, containing a well-defined
center of rotation,
as seen in figure \ref{F:stages}(b) 
and the associated boundary layer of negative vorticity.
In this stage the vortex center grows, but does not
satisfy scaling laws. Luchini and Tognaccini refer to 
it as the viscous stage.
In the third stage, the self-similar stage, 
loosely represented by figure \ref{F:stages}(c), 
the vortex center grows closely satisfying the self-similar 
scaling for inviscid separation in the absence of
a plate length scale. The observed scaling behaviour is the 
subject of the next section.
In the last stage, the ejection stage,
illustrated in figure \ref{F:stages}(d),
the vortex departs from the self-similar
growth, and the finite plate length 
noticeably affects the flow.


%
%
%
%

  \begin{figure*}
  \includegraphics[width=0.32\textwidth]{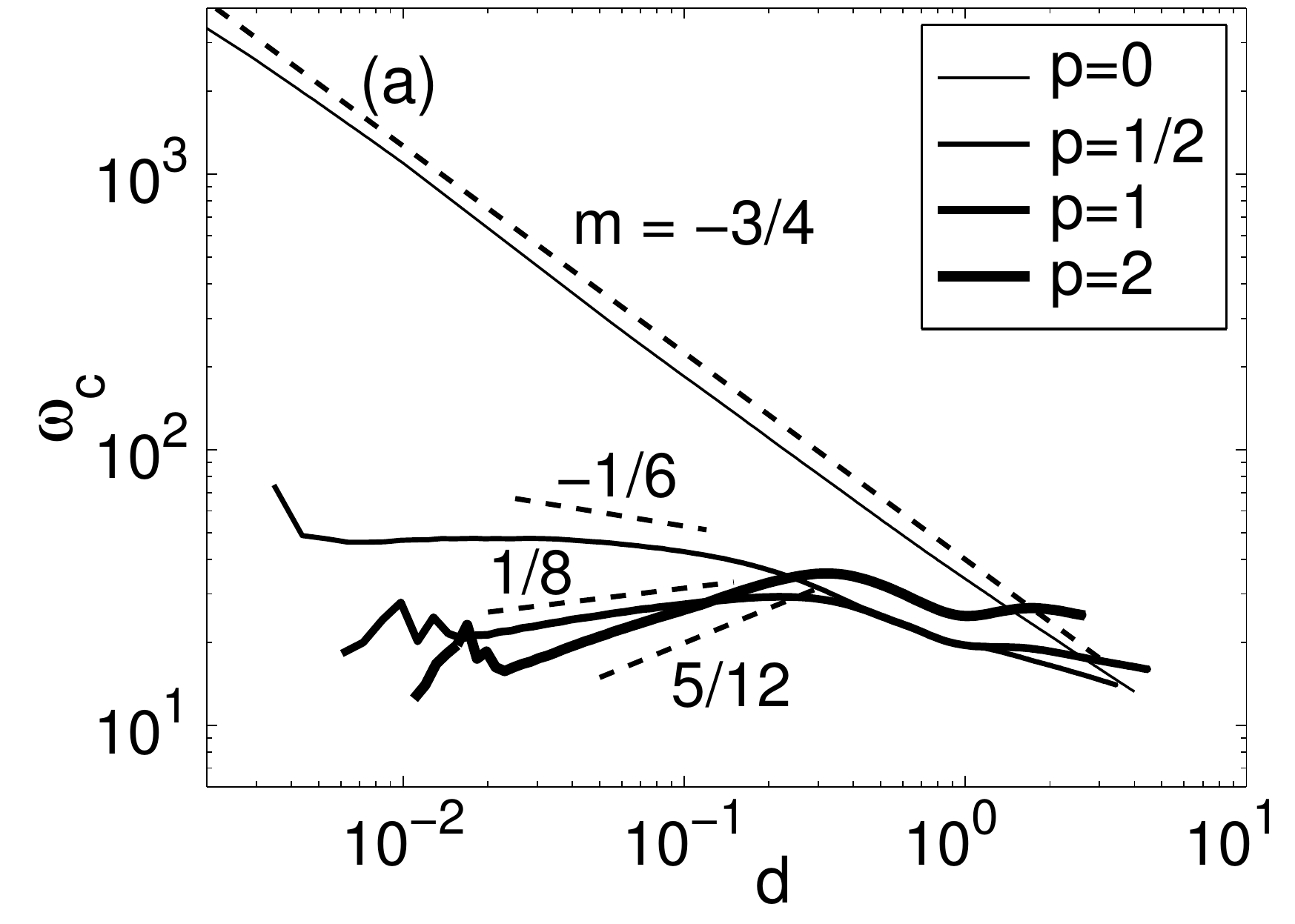}
  \includegraphics[width=0.32\textwidth]{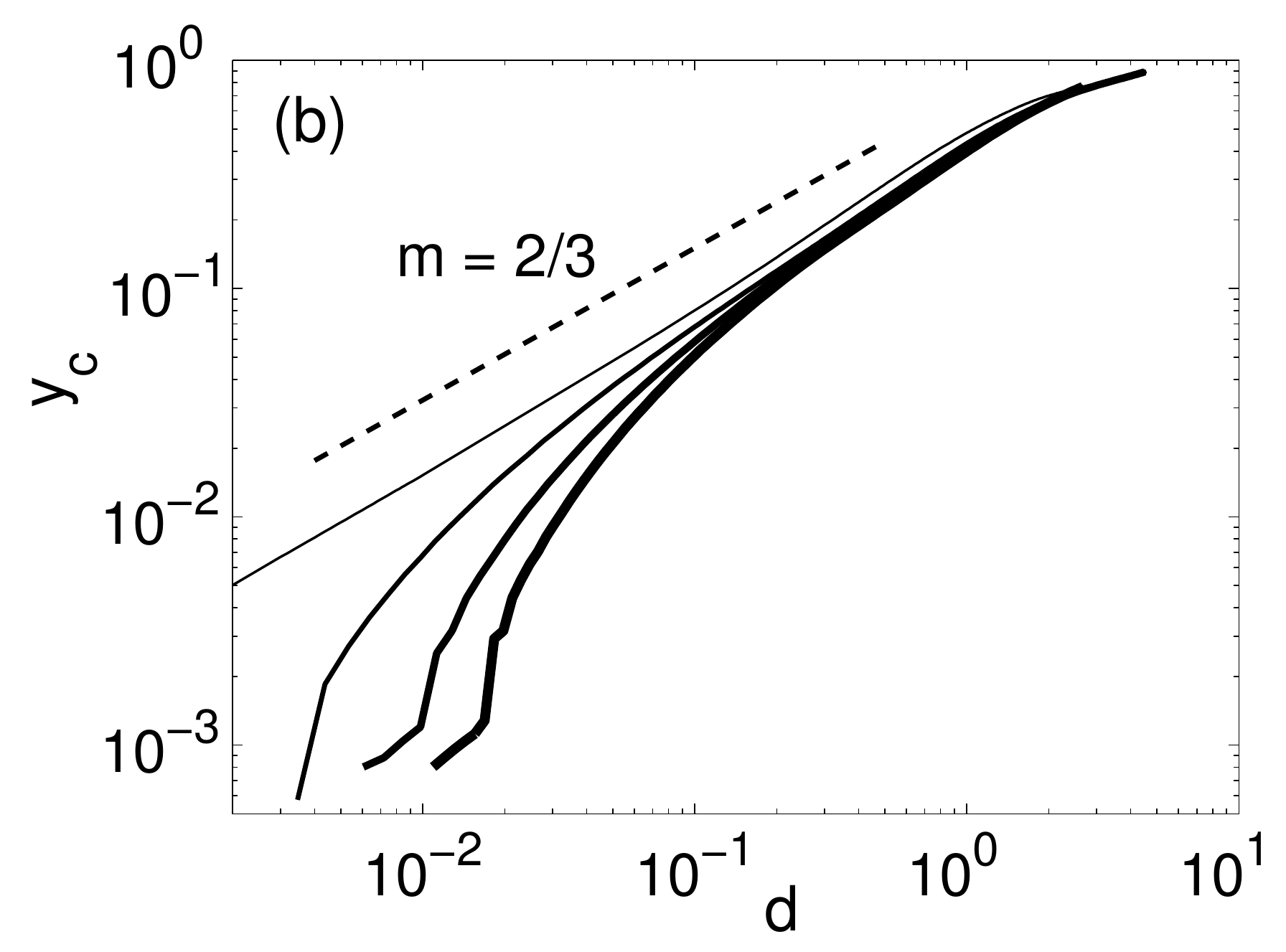}
  \includegraphics[width=0.32\textwidth]{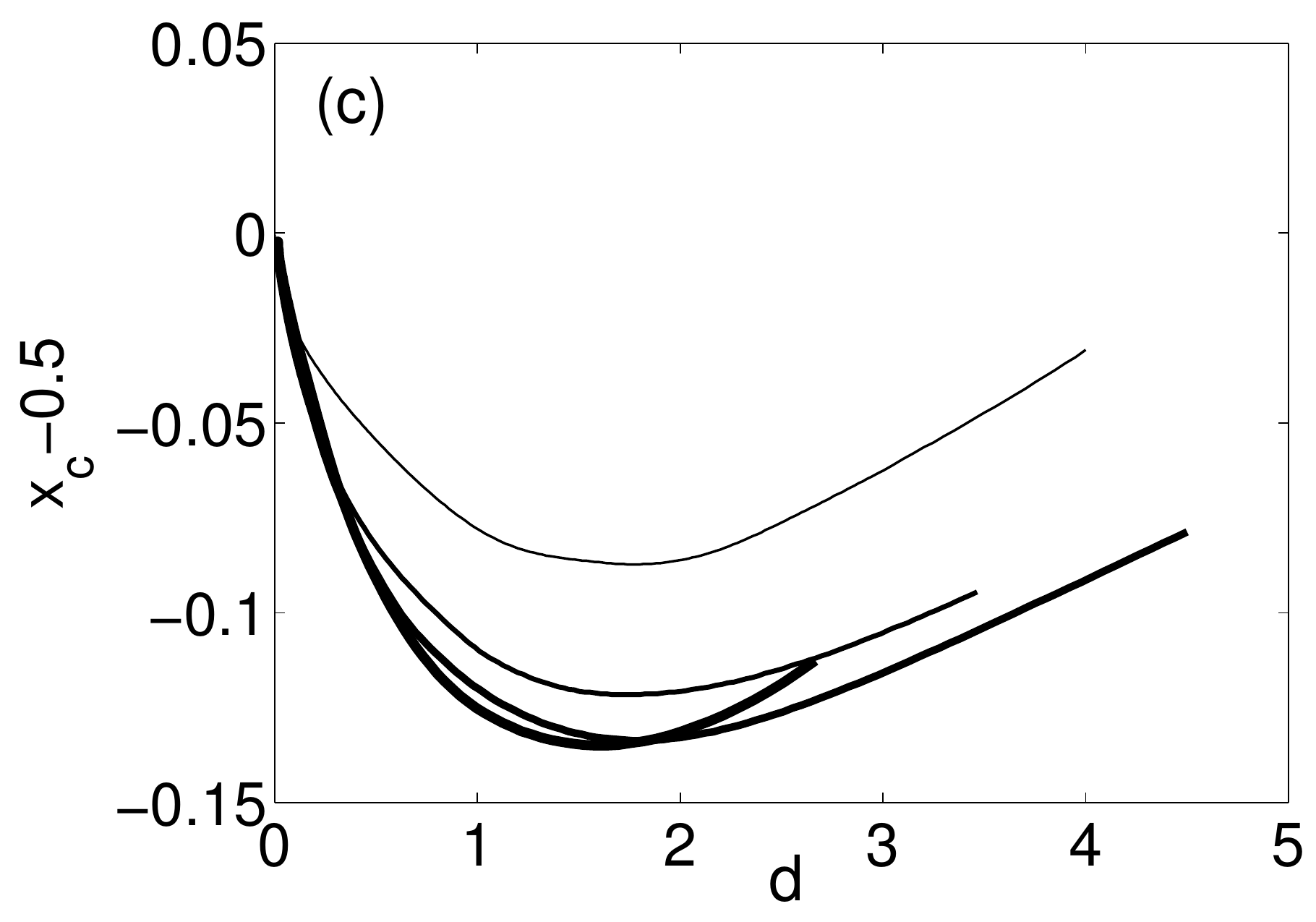}
\vskip-4.2truecm  
\hbox{ } \hskip12.1truecm
\includegraphics[width=0.12\textwidth]{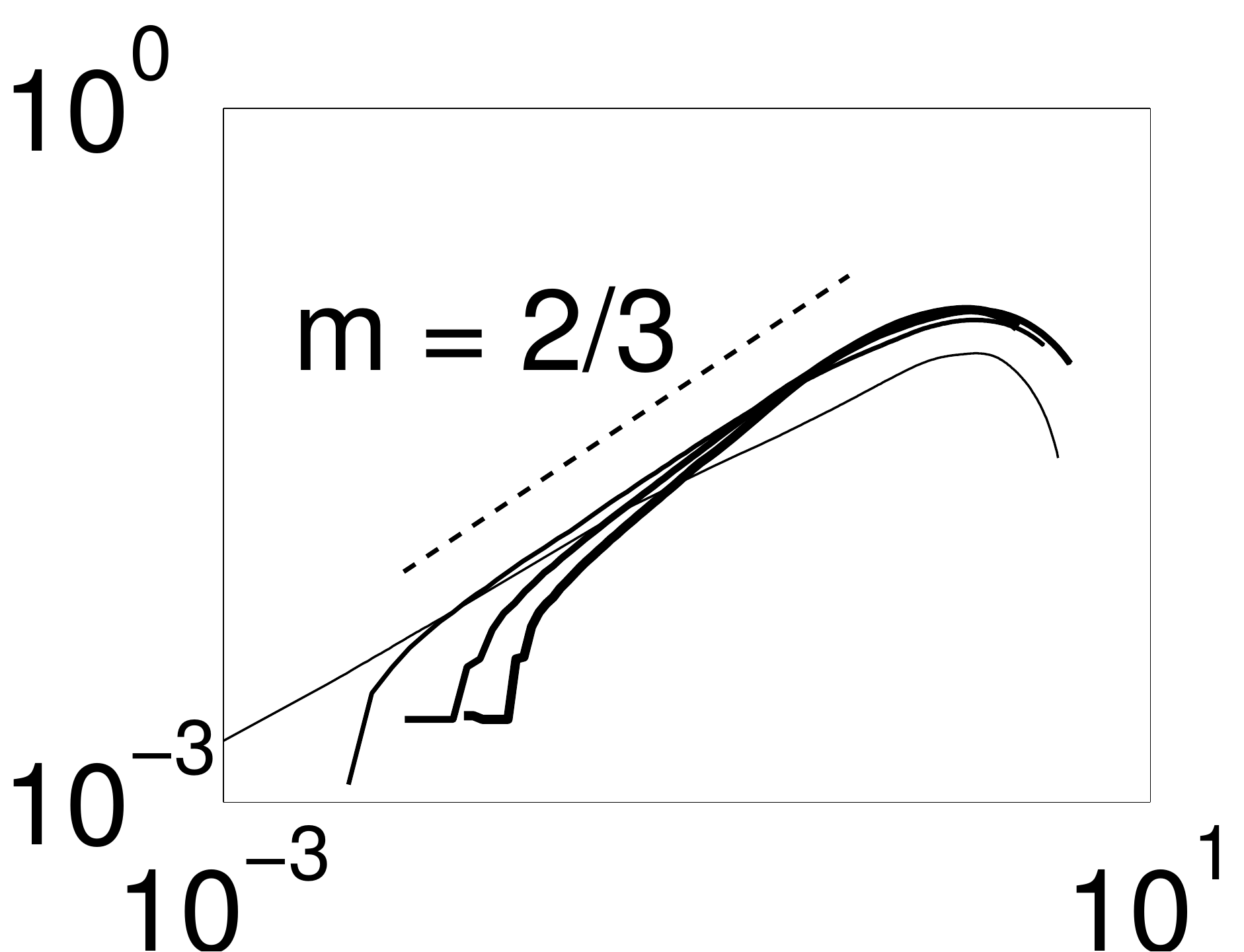}
\\
\vskip2.1truecm  
  \includegraphics[width=0.32\textwidth]{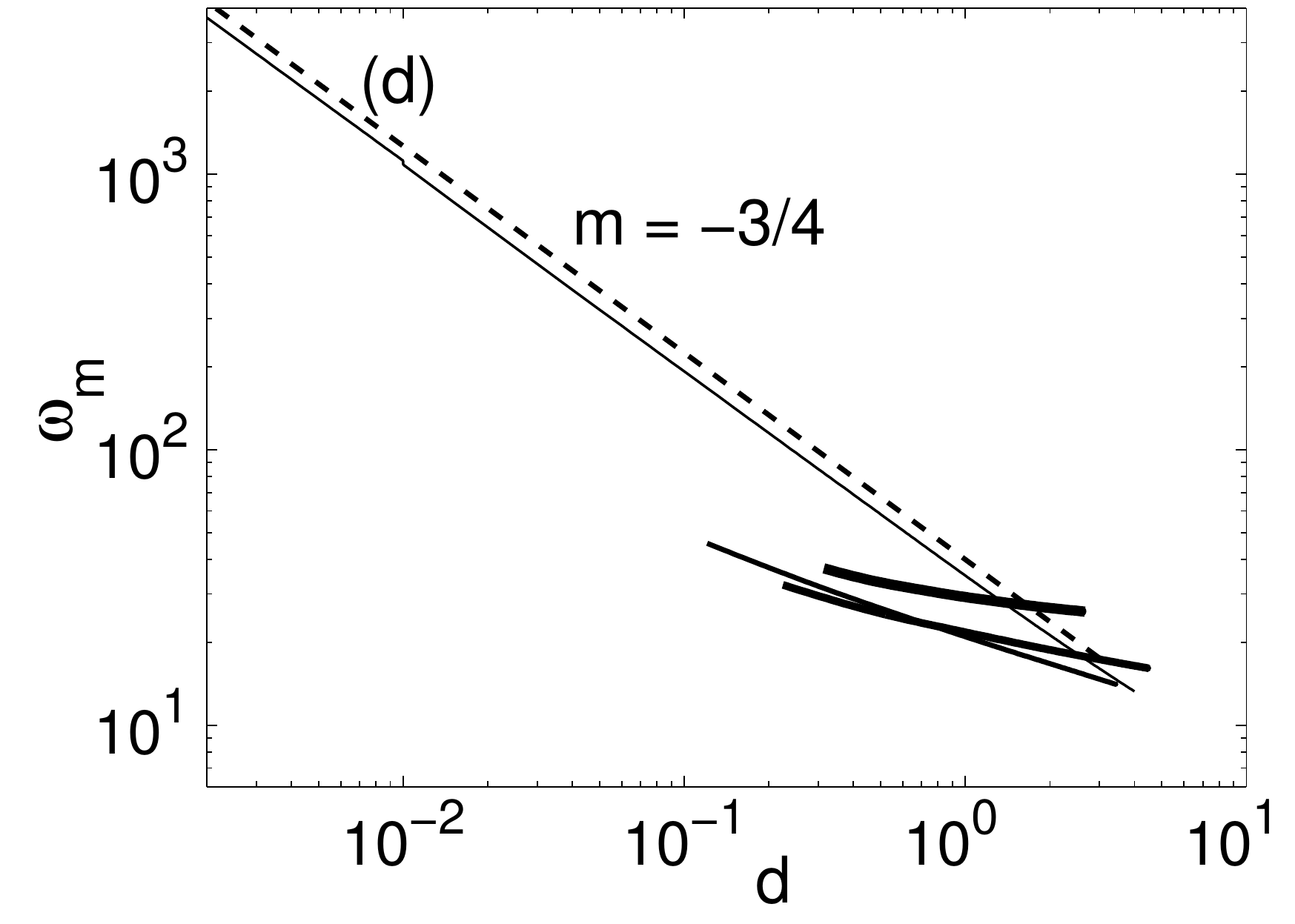}
  \includegraphics[width=0.32\textwidth]{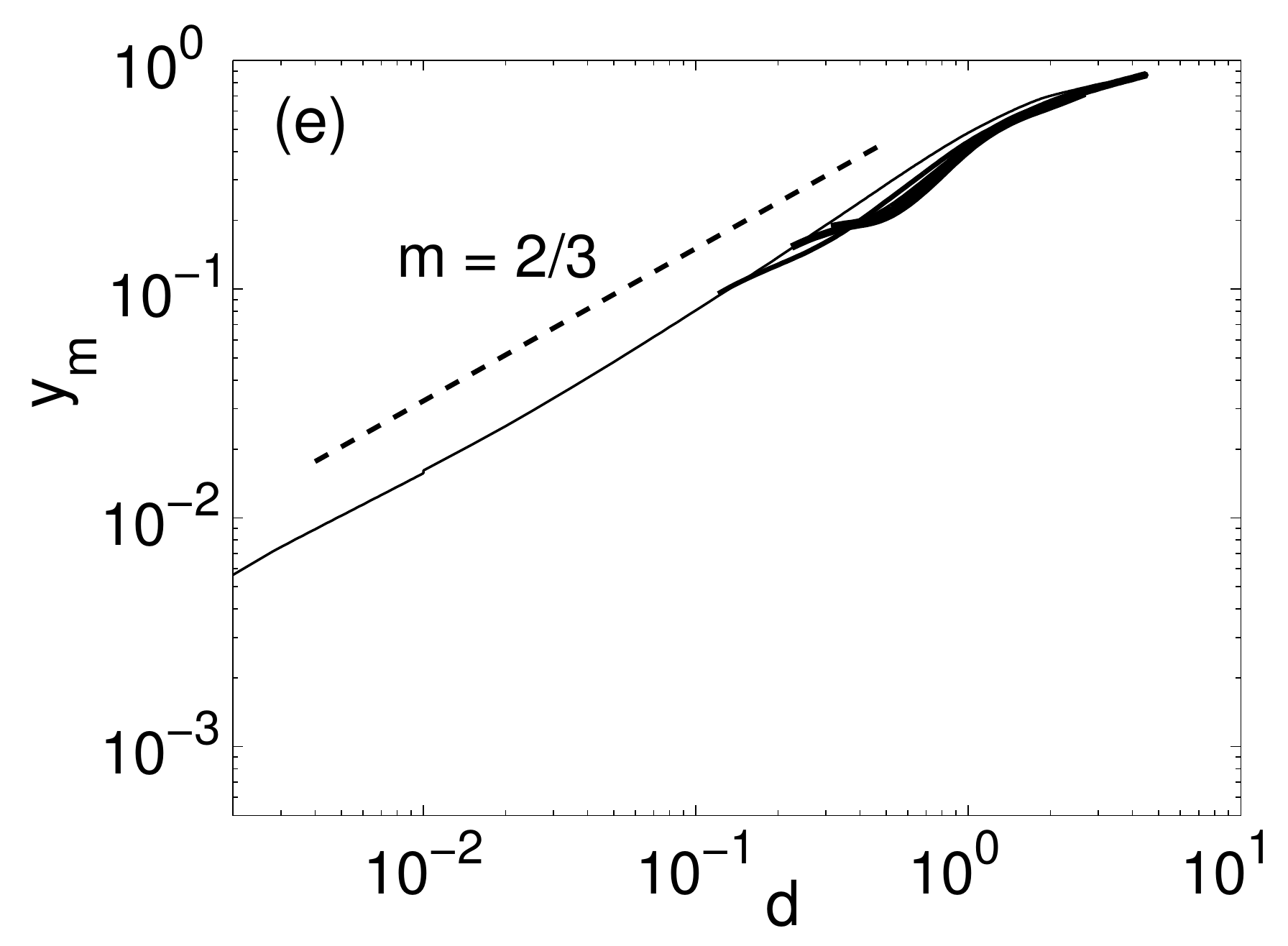}
  \includegraphics[width=0.32\textwidth]{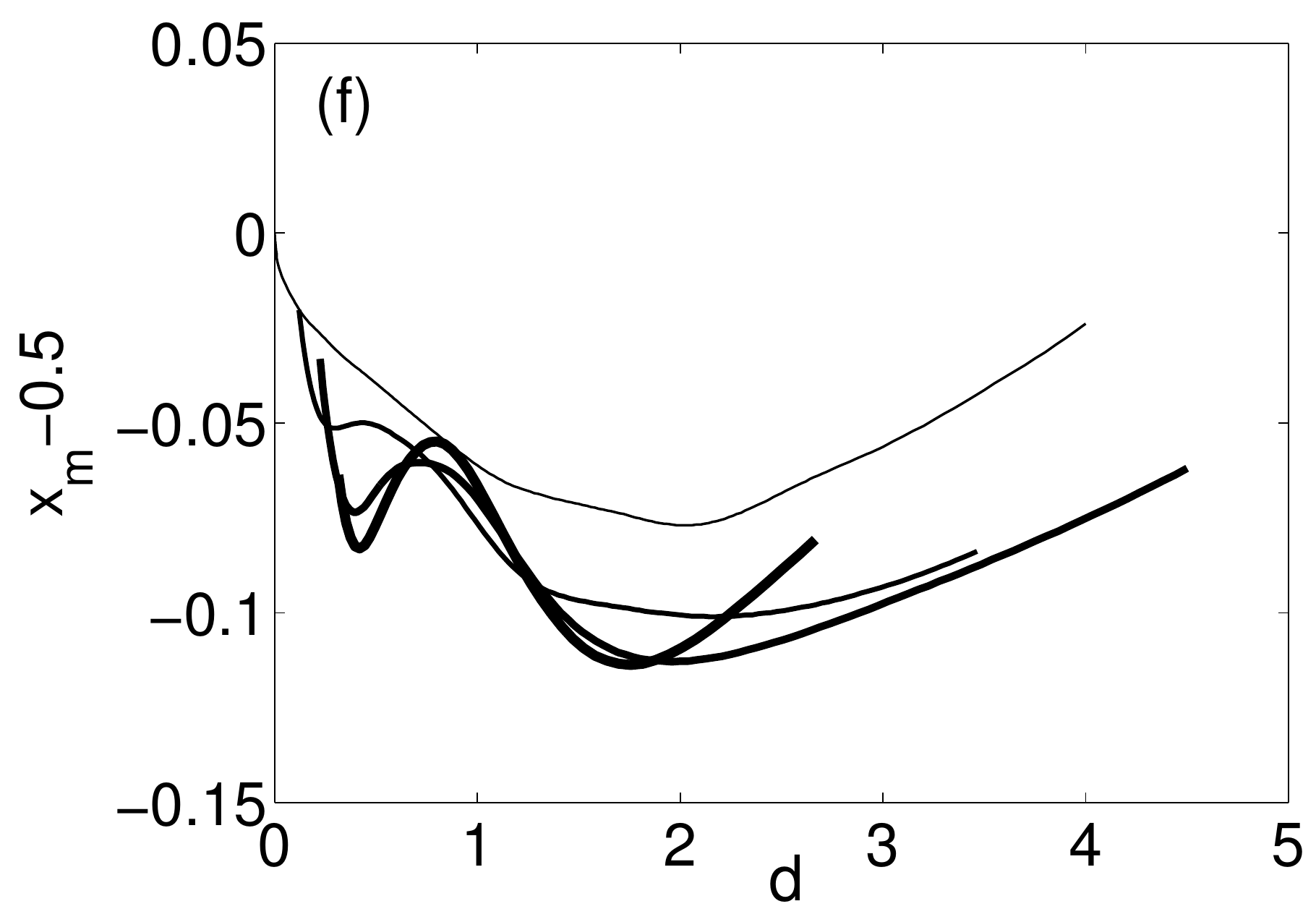}
  \caption{
(a-c) Vorticity and coordinates of the rotation center, \textit{vs.} displacement $d$.
(d-f) Vorticity and coordinates of the core vorticity maximum center, \textit{vs.} displacement $d$.
Results are shown for $p=$0, 1/2, 1, 2, as indicated in the legend in (a).
The dashed lines have the indicated slopes.}
  \label{F:scaling_core}
  \end{figure*}

It is interesting to note that the Rayleigh stage
is not observed for the impulsively started flow, $p=0$. 
In that case (see also \cite{xunitsche14}),
the recirculation region near the plate tip and the associated negative vorticity
appear within the first 10-15 timesteps, that is, in time $O(\Delta t)$.
Figure \ref{F:tr}
plots the duration $T_R$ of 
the Rayleigh stage, defined by the time at which the 
recirculation region and negative vorticity first appear,
as a function of $p$.
The results where computed using
two resolutions, $h=1/1280$ and $1/640$, as indicated, 
for $p=1/8,1/4,1/2,1,2$. 
While the convergence is slower for smaller values of $p$,
the interpolating curves appear to converge to 
\begin{equation}
 T_R \approx 0.15p^{3}.
\end{equation}
Thus, $T_R$ grows as $p$ increases, and vanishes for $p=0$,
consistent with our observations. 

\subsection{Vortex core trajectory and vorticity}
\label{trajectory}

At $t=T_R$, a region of recirculating flow forms with
an associated center of rotation, which 
can be used to define the center of the starting vortex.
Alternatively, the vortex center can be defined as the
position of the vorticity maximum near its center. 
However, as can be seen in figure \ref{F:stages}(b), 
at early times when the recirculation region is already
well established, no local vorticity maximum 
has yet formed
within the starting vortex.
At these early times, the vorticity grows along a curved ridge starting
at its maximum value at the plate tip, but the ridge
does not develop a local maximum along it until much later. 
Once the local vorticity maximum appears,
as in figure \ref{F:stages}(c,d), it is not at the
same location as the center of rotation.
We denote the positions of 
the center of rotation and the vorticity maximum
by $(x_c,y_c)$ and $(x_m,y_m)$, respectively,
and the corresponding vorticity at those points by $\omega_c$ and $\omega_m$
(see figure \ref{F:stages}(d).

Figure \ref{F:comptraj} compares the trajectory of the 
rotation center (solid) and the core vorticity maximum (dashed)
for $p=0, 1/2 ,1, 2$, computed for displacements $d\in[0, 3]$.
The figure shows the trajectories on a one-to-one scale,
showing that the vortex travels much faster in the vertical 
direction away from the plate than in the horizontal direction. 
The vorticity maximum lies always to the right of the rotation center. 
The figure shows that as $p$ increases, the 
vorticity maximum appears later and at a further distance from
the plate tip.
Furthermore, as $p$ increases, the difference between the
two points increases. 
Finally, while the rotation center travels on a monotonic path
inward from the plate tip for most of the interval shown,
the vorticity maximum oscillates as it travels downstream.
It is possible that such oscillations in the core vorticity
is partially responsible for oscillations along the separated shear
layer often observed in laboratory experiments and in computations 
\cite{koushiels96,luchini02,taneda71,wang99}.
However, this issue remains to be investigated. 

Figure \ref{F:scaling_core} plots the
vortex core vorticity and coordinates as a function of the
displacement $d$, on a logarithmic scale, in order to reveal their scaling
behaviour. 
The top row, figures \ref{F:scaling_core}(a,b,c), shows the results 
$\omega_c,y_c, x_c$ for the rotation center, which is the first to form.
The bottom row, figures \ref{F:scaling_core}(d,e,f), shows the results 
$\omega_m,y_m, x_m$ for the vorticity maximum.

We first discuss the results for core vorticity shown in 
figures (a,d).
As noted in \cite{xunitsche14}, for $p=0$ the vorticity closely follows
the viscous scaling found in flow past a semi-infinite plate, 
over several decades in time.
Consider flow past a semi-infinite plate driven by $\psihat_{\infty}=A\that^p\rhat^{1/2}\cos(\theta/2)$, where $\rhat,\theta$ are the polar coordinates of a point
with origin at the plate tip, and $A$ is a dimensional constant.
Due to the absence of a plate length scale,
it follows from
dimensional analysis that the dimensional flow streamfunction has the
form
\begin{equation}
\psihat(\xhat,\yhat)=A\that^p(\nu \that)^{1/4}f({\xhat\over\sqrt{\nu \that}},{\yhat\over\sqrt{\nu \that}})~.
\label{E:dimanal}
\end{equation}
In our case, this scaling can be expected to be a good 
approximation at early times, as long as the vortex size 
is small relative to the finite plate length.
By taking second derivatives, one finds that the corresponding
vorticity scales as 
$\what\sim A\that^p(\nu \that)^{-3/4}$.
In our nondimensional variables, using $A=L^{1/2}a$, 
this implies that
\begin{equation}
\omega\sim t^p\Big({t\over Re}\Big)^{-3/4}
\sim  d^{p-3/4\over p+1}
=d^{\alpha}~,
\label{E:omegascaling}
\end{equation}
where 
$\alpha=-3/4$, $-1/6$, $1/8$, $5/12$ for $p=0,1/2,1,2$, respectively.
As shown in figure \ref{F:scaling_core}(a,d),
for $p=0$ this scaling is observed over a large range of times,
both in $\omega_c$ and $\omega_m$.  
For $p>0$, the scaling is observed in $\omega_c$, but over 
a much smaller time interval, starting after an initial transition
region and ending aproximately at $d=0.2$. 
Figure \ref{F:scaling_core}(d) shows that for $p>0$, 
the vorticity maximum $\omega_m$ has not even yet formed
during those times.  It appears only
around $d=0.2$ and does not follow the scaling (\ref{E:omegascaling}).

Figures \ref{F:scaling_core}(b,e) plot the vertical 
displacements $y_c,y_m$ of the rotation center 
and the vorticity maximum, respectively, 
as a function of the plate displacement $d$, respectively. 
For $p=0$, both variables closely satisfy
\begin{equation}
 y_c,y_m \sim d^{2/3}
\label{E:yscale}
\end{equation}
over several decades in time, until about $d=1$. 
For $p>0$, the data shows that $y_c$ approximates the same 
scaling quite well for roughly $d>0.2$, 
after an initial transition period.  
As already noted, the values of $y_m$
do not exist during this
transition period. 
They are in fair agreement with $y_c$ afterwards. 
as shown in figure \ref{F:scaling_core}(e). 

Figures \ref{F:scaling_core}(c,f) plot the horizontal 
displacement of the vortex center from the plate tip,
$x_c-0.5$ and $x_m-0.5$, respectively, as a function of the displacement $d$.
Figure (c) shows that the center of rotation first moves monotonically
towards the axis, as could also be seen in figure \ref{F:comptraj}.
For larger values of $p$, it moves further to the left. 
Approximately around $d=1.5$, for all $p$, the vortex turns around and moves 
outwards. 
The inset in figure (c) plots $|x_c-0.5|$ on a logarithmic scale,
and shows that during the inward motion, the scaling
\begin{equation}
 |x_c-0.5| \sim d^{2/3}~, 
\label{E:xscale}
\end{equation}
is approximately satisfied. For $p>0$, it is satisfied well 
after an initial transition period.
The data for $p=0$ appears to have a slightly smaller slope,
between $2/3$ and $1/2$.
The reason for this apparent jump between $p=0$ and $p=0.5$ is not 
quite clear at this point, and needs to be investigated further. 
The data for $x_m$, shown in figure \ref{F:scaling_core}(f),
shows that the vorticity maximum oscillates as it moves
inward towards the axis, before it moves back out,
as was already observed in figure \ref{F:comptraj}.
The oscillation amplitude increases as $p$ increases.

Note that the observed scaling in equations 
(\ref{E:yscale},\ref{E:xscale}) 
is not the one that follows for viscous
flow, from the argument
leading to (\ref{E:dimanal}). Instead, it is the scaling found for 
inviscid flow in the absence of a length scale, also based on dimensional
analysis. This case was considered 
by Pullin \cite{pullin78}, who derives the self-similar scaling 
for the vortex sheet separation and spiral roll-up
at the edge of a semi-infinite plate
driven by accelerating background flow,
and computes the time-independent self-similar shape
using an iterative scheme. His results show that 
the coordinates of the spiral center satisfy
\begin{equation}
\xhat_c+i\yhat_c=\Omega_0\Big({A\that^{\, 1+p}\over 1+p}\Big)^{2/3} 
\label{E:pullinxy}
\end{equation}
where the complex number $\Omega_0$ depends little on $p$.
In the dimensionless variables used here, and again making the 
correspondence $A=L^{1/2}a$, equation (\ref{E:pullinxy}) is 
equivalent to equations (\ref{E:yscale},\ref{E:xscale}).


  \begin{figure}
  \centering
\includegraphics[width=0.42\textwidth]{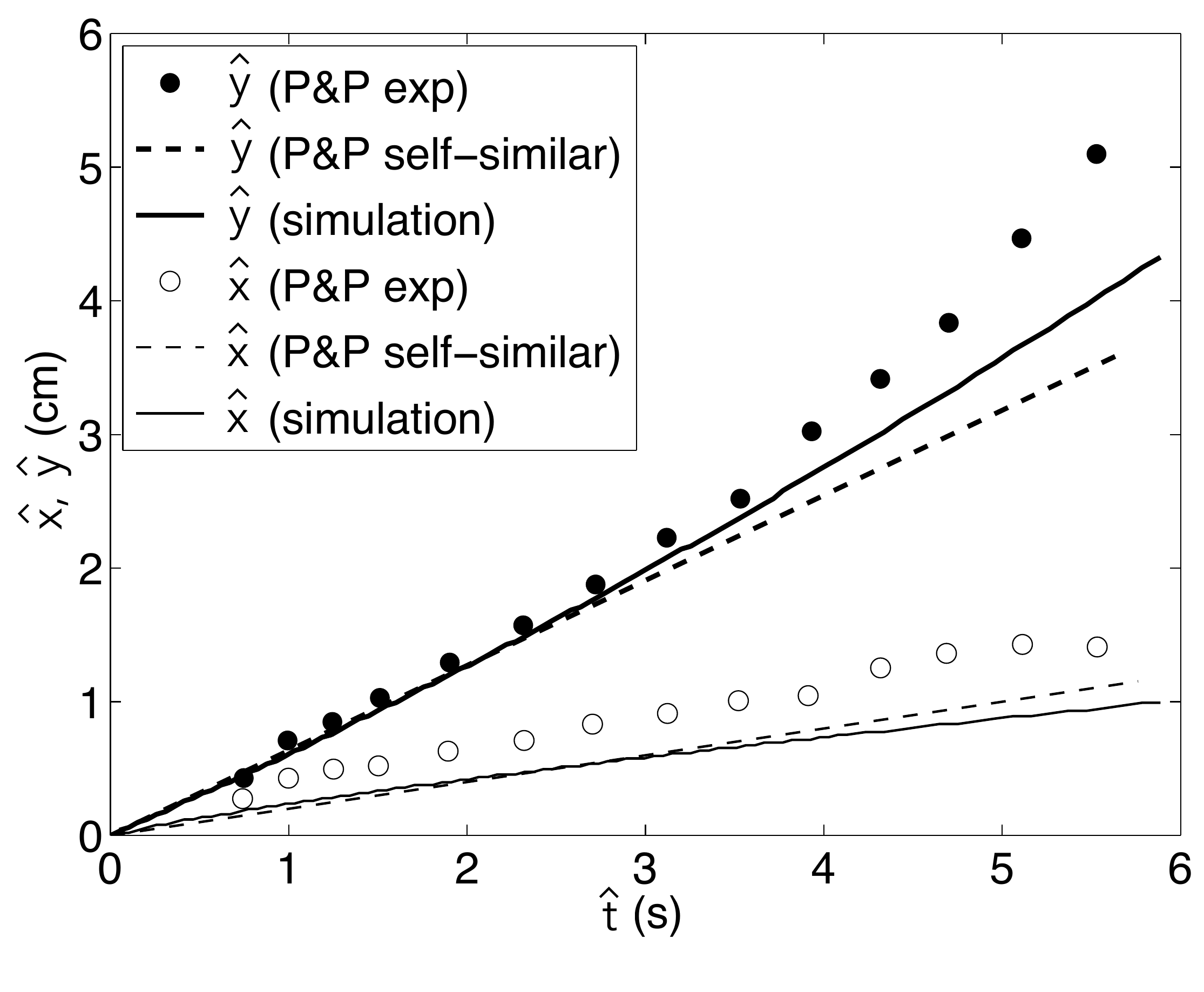}
  \caption{
Comparison of dimensional vortex core coordinates $(\xhat,\yhat)$ 
relative to the plate tip, for $p=0.45$. Experimental results of Pullin 
\& Perry \cite{pullinperry80} ($Re=6621$),
inviscid similarity theory results\cite{pullin78,pullinperry80},
and numerical results for the rotation center ($Re= 6000$)
are shown.
}
  \label{F:pullin_core}
  \end{figure}


  \begin{figure*}
  \centering
  \includegraphics[width=0.315\textwidth]{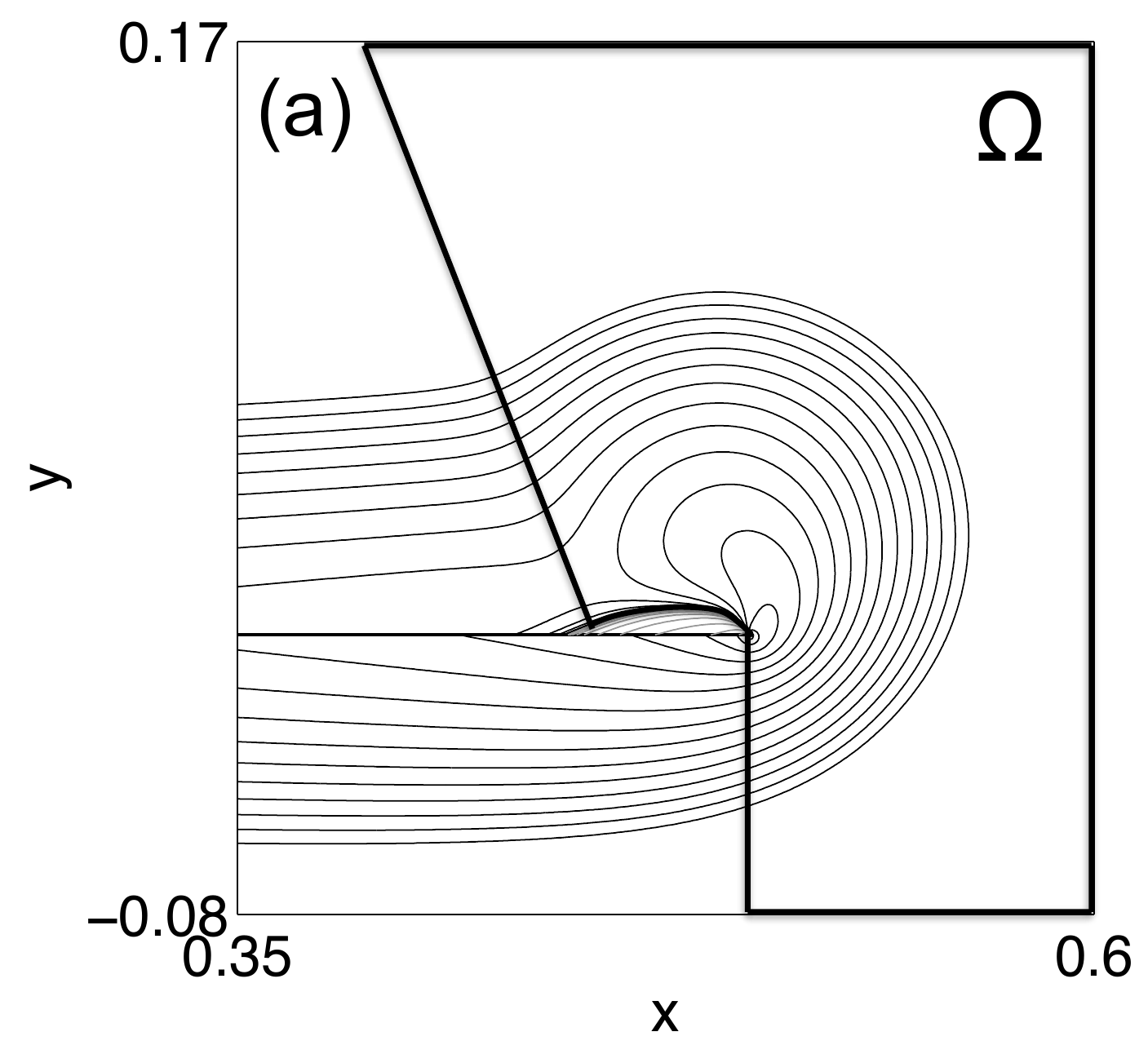}\hskip10pt
  \includegraphics[width=0.304\textwidth]{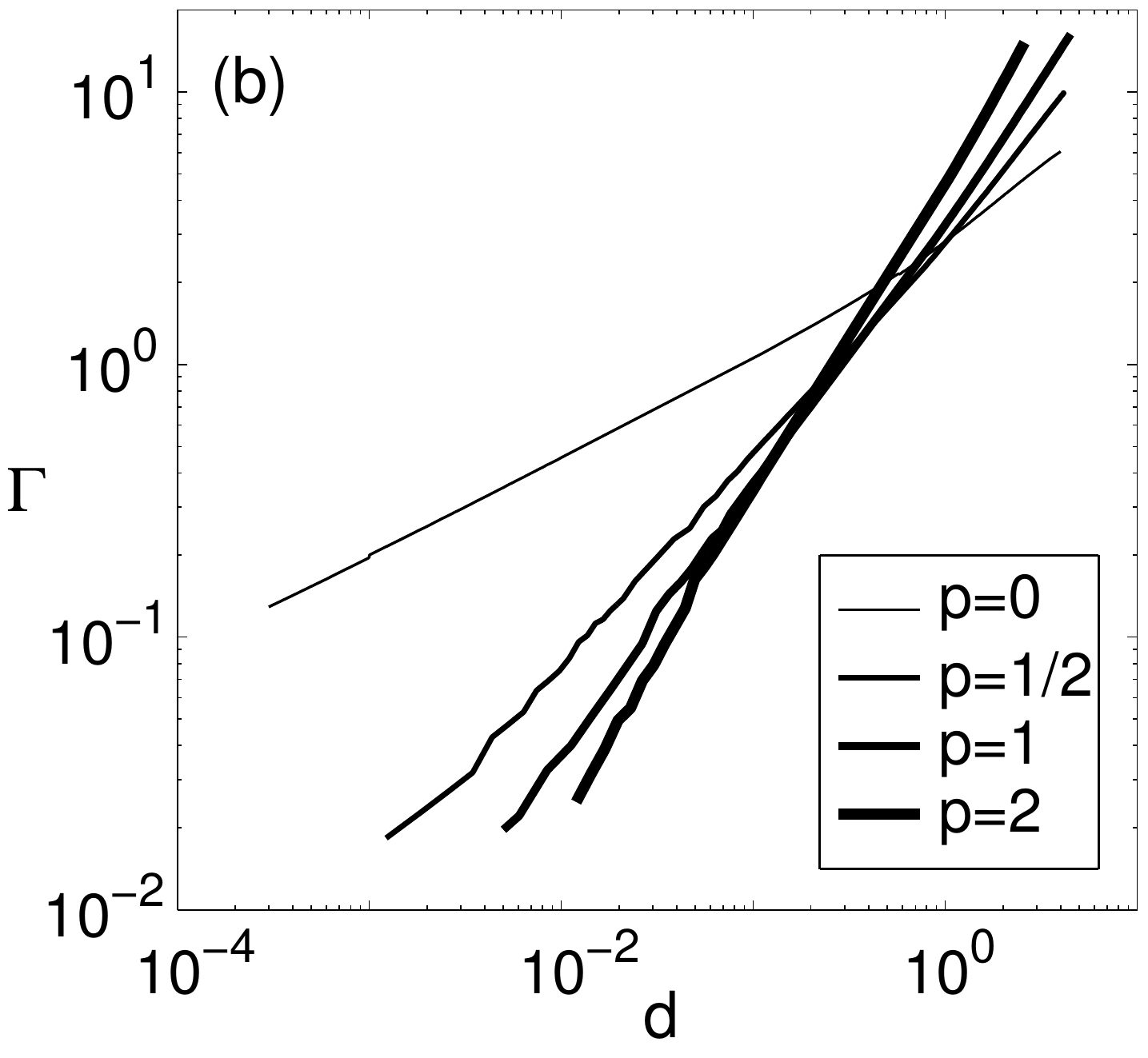}\hskip10pt
  \includegraphics[width=0.32\textwidth]{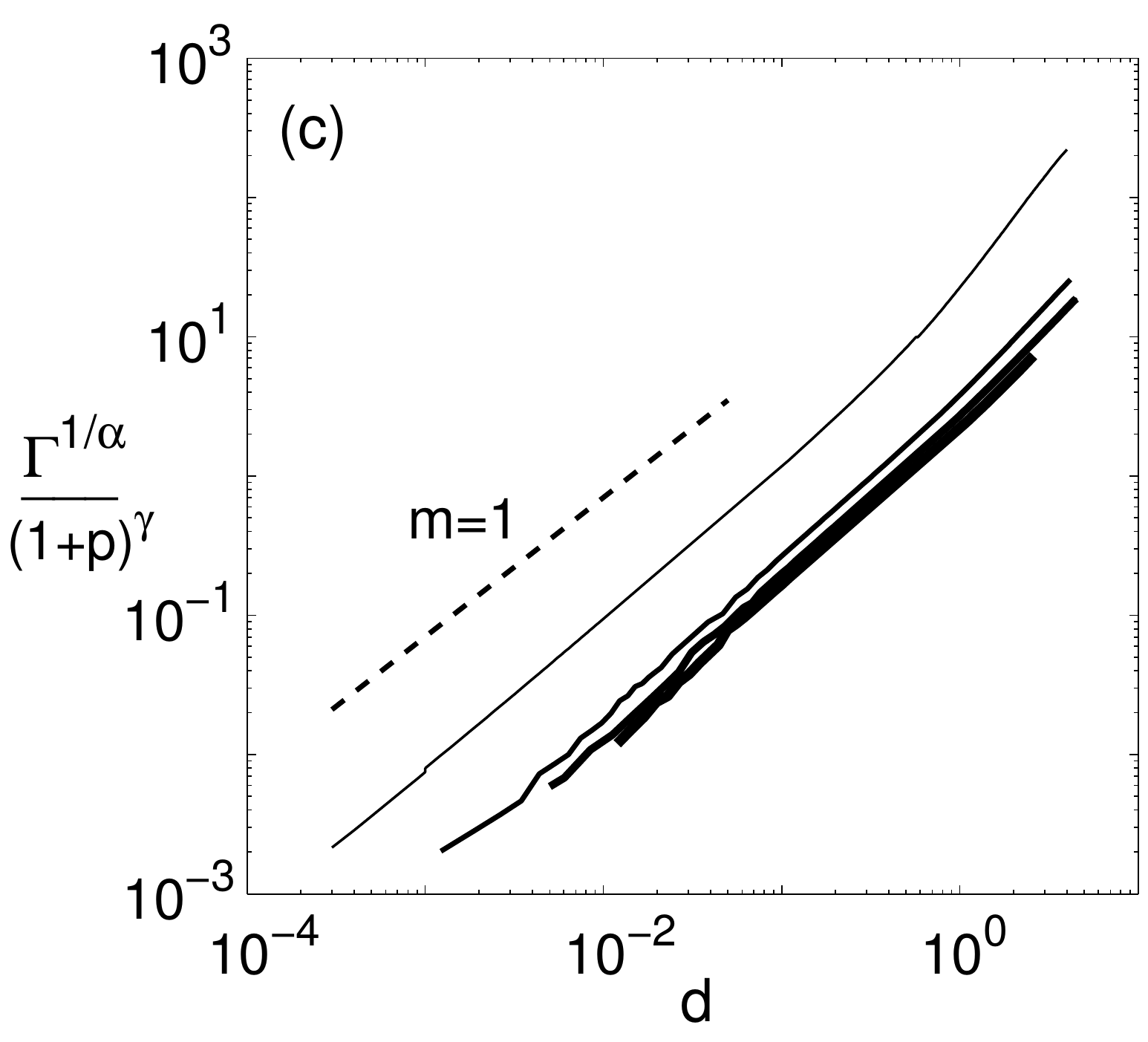}
  \caption{(a) Sketch showing domain $\Omega$ used to define 
shed circulation $\Gamma$, following \cite{xunitsche14}. 
(b) $\Gamma$ $vs$ displacement $d$. 
(c) $\Gamma^{1/\alpha}/(1+p)^{\gamma}$ $vs$ $d$. 
where $\alpha={1+4p\over 3(1+p)}$, $\gamma={3p\over 1+4p}$.
Results are shown for $p=0$, $1/2$, 1 and 2, as indicated in (b). 
The dashed line in (c) has the indicated slope.}
  \label{F:gam}
  \end{figure*}

Next we compare
not only the scaling behaviour, but the 
actual values of the core coordinates
with experimental data of Pullin \& Perry \cite{pullinperry80} (P\&P)
and with similarity theory results given therein.
Figure \ref{F:pullin_core} plots 
the dimensional coordinates $(\xhat,\yhat)$ of the vortex
center relative to the plate tip as a function of
dimensional time $\that$.
Consistent with the rest of this paper, $\xhat$ 
refers to the vortex displacement parallel to the plate, 
while $\yhat$ refers to the displacement normal to the plate.
The experimental measurements obtained by P\&P
(closed and open circles)
represent the spiral center of the experimentally observed streakline,
for flow past a wedge with $\beta=5^{\circ}$. 
The similarity theory results obtained by Pullin\cite{pullin78} 
(thick and thin dashed curves)
also correspond to flow past a wedge with $\beta=5^o$. 
The computed results (thick and thin solid curves)
denote the dimensional position of the rotation center 
relative to the plate tip, for a plate with
$\beta=0$, $Re=6000$, $p=0.45$. 
They are dimensionalized appropriately, 
\begin{equation}
\xhat=L(0.5-x_c)\,,~\yhat=Ly_c
\end{equation}
and shown at dimensional times $\that=Tt$. 
The times $\that\in[0,6]$ shown in the figure correspond
to nondimensional times $t\in[0,0.58]$, or displacements $d\in[0,0.31]$.
At these times the vortex center still travels towards the axis,
with decreasing values of $x_c$ (see figure 9c), that is,
increasing values of $\xhat$.

Figure \ref{F:pullin_core} shows that 
the computed values for the component $\yhat$ normal to the plate
is in excellent agreement with both the experimental data and 
similarity theory, for relatively long times $\that\le 3$ ($d\le 0.1$).
After this time all three values begin to differ, indicating that 
the finite plate length and possibly the wedge angle and the physical
wall along the centerline begin to affect the results. 
The computed values remain in between the experimental and the similarity
theory results.
%
The computed values for the component $\xhat$ 
tangent to the plate is in less agreement with the experimental 
data. The experiments show larger deviation from the tip. 
However, the computed values of $\xhat$, corresponding to $\beta=0^o$,
are in surprisingly good agreement with similarity theory results
for $\beta=5^o$.

\subsection{Vortex circulation}\label{circulation}

This section presents the shed circulation $\Gamma$ as 
a function of $p$. No vorticity 
has separated 
for $t<t_R$, during which there is a Rayleigh boundary around the whole plate,
as shown in figure \ref{F:stages}(a). 
After this time, a region of recirculating flow has
formed near the tip which one can associated with
a starting vortex. However, the corresponding vorticity is
embedded in the boundary layer vorticity, and it
is not clear a priori how to define shed vorticity.
In order to distinguish separated from attached vorticity,
we use the fact that for $t>t_R$, the vorticity contours above the 
plate develop a point of maximal curvature.
These high curvature points closely follow a slant line
to the left of the vortex, as shown in figure \ref{F:gam}(a). 
We follow our earlier work$\cite{xunitsche14,nitschexu14}$ and define the 
separated vorticity to be that
enclosed
in the region $\Omega$ shown in figure \ref{F:gam}(a). 
It includes all vorticity to the right of the tip,
and all positive vorticity to the left of the slant line.
The negative vorticity attached to the wall is excluded,
although it is convected into $\Omega$ at later times,
when it enters $x\ge 0.5$, 
such as in figures \ref{F:stages}(c,d).
This definition has a continuous and natural extension 
to later times, see \cite{xunitsche14}.
The shed circulation is defined to be
\begin{equation}
 \Gamma(t) = \int_{\Omega(t)}\omega(\cdot,t) dA~.
\end{equation}

Figure \ref{F:gam}(b) plots the circulation $\Gamma$,
computed with this definition,
as a function of the displacement $d$, for $p=0,1/2,1,2$, as indicated.
Each curve begins at the value of $d$ corresponding to $t=t_R$.
For each $p$, the curves closely follow a straight line
in the logarithmic scale shown, 
indicating a power law behaviour $\Gamma\sim d^{\beta}$. 
The slopes $\beta$ increase
with increasing $p$. 
They are in fact in close agreement with inviscid similarity theory.
Pullin\cite{pullin78} shows that for the self-similar
inviscid vortex sheet separation at the edge of 
a semi-infinite plate, driven by a power law background flow,
the separated vortex sheet circulation $\Gamma_{vsh}$ satisfies
%
\begin{subequations}
\begin{equation}
\Gamma_{vsh}\approx J {t^{1+4p\over 3}\over (1+p)^{1/2}}
\end{equation}
where $J$ is only weakly dependent on $p$. This equation can be rewritten as
\begin{equation}
{\Gamma_{vsh}^{1/\alpha}\over (1+p)^\gamma }\approx J^{1/\alpha}d 
\label{E:selfsim_circ}
\end{equation}
\end{subequations}
where $\alpha={1+4p\over 3(1+p)}$
and $\gamma=1-{1\over 3\alpha}={3p\over 1+4p}$. 
To compare the present results for viscous vortex separation
with the inviscid similarity theory, 
figure \ref{F:gam}(c) plots  $\Gamma^{1/\alpha}/(1+p)^{\gamma}$,
vs the displacement $d$.
The curves for all $p$ are almost parallel and have
slope approximately equal to 1, showing that 
the power law scaling (\ref{E:selfsim_circ}) 
is closely satisfied,
with 
\begin{equation}
\Gamma\sim {{1\over (1+p)^{1/2}}}t^{1+4p\over 3}~.
\end{equation}
However the constant of proportionality depends significantly
on $p$ and increases by a factor of almost 10 as $p$ decreases 
from $p=2$ to $p=0$.
The dependence on $p$ increases as $p$ decreases,
with a large difference between $p=0$ and $p=1/2$, while the
results for $p\ge 1/2$ change little.

\subsection{Characteristic sizes}\label{boundarylayer}

\begin{figure}
  \centering
  \includegraphics[width=0.43\textwidth]{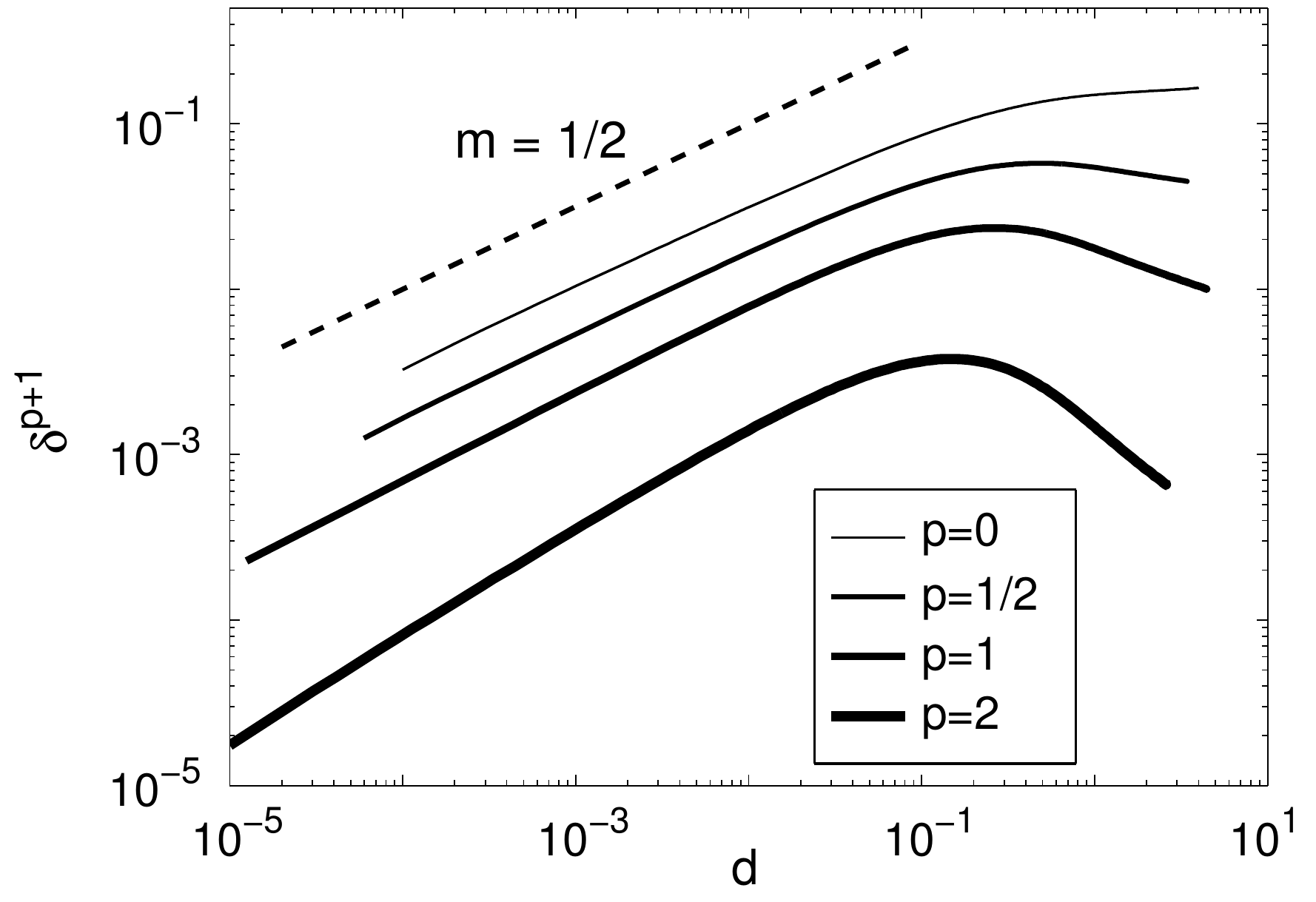} ~ ~
  \caption{Boundary layer thickness $\delta$
of the region with $\omega\ge 2^{-12}$ at $x=0.25$, below the plate.
The quantity $\delta^{p+1}$ is plotted $vs$ displacement $d$, 
for $p$=0, 1/2, 1 and 2, as indicated.
The dashed line has the indicated slope.}
  \label{F:blayer}
  \end{figure}

To conclude, this section presents two characteristic
sizes of the flow: the boundary layer thickness $\delta$ and the
size of the recirculation region $s_{\Psi}$.
Figure \ref{F:blayer} plots $\delta^{p+1}$ $vs$ displacement $d$,
where $\delta$ is the vertical thickness
of the region below the plate with $\omega \ge 2^{-12}$,
at $x=$0.25. 
The thickness decreases noticeably as $p$ increases.
The figure shows that 
\begin{equation}
 \delta^{p+1} \sim d^{\frac{1}{2}}\quad\hbox{or}\quad
 \delta \sim t^{\frac{1}{2}}
\end{equation}
for approximately $d\in[0,0.1-0.4]$. The scaling holds longer
for smaller values of $p$. 
The thickness at other
values of $x$ is qualitatively similar. 
Thus, the boundary layer
grows as expected, satisfying the 
viscous scaling of equation \ref{E:dimanal}.

\begin{figure}[ht!]
  \centering
  \includegraphics[width=0.43\textwidth]{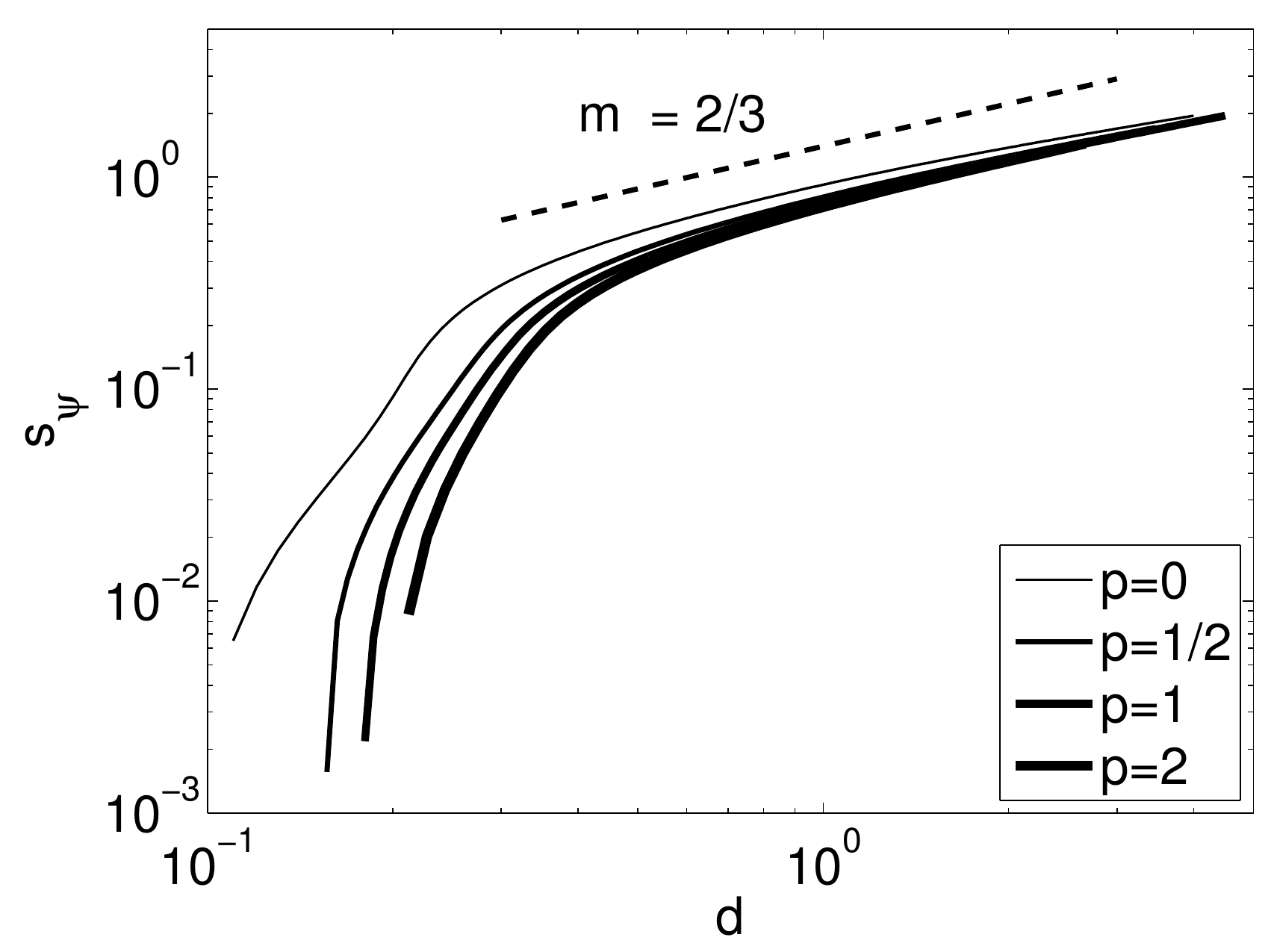}
  \caption{Vortex size $s_\psi$ vs displacement $d$, 
for $p$=0, 1/2, 1 and 2, as indicated.
The dashed line has the indicated slope.}
  \label{F:size}
  \end{figure}

The size of the vortex pair separating at the edge of a finite
plate has previously been reported in terms of the quantity
$s_\psi$, defined as shown in figure \ref{F:stages}(d) to be 
the height of the recirculation region on the axis of symmetry
along the middle of the plate.
Note that this quantity is defined only after
the recirculation region has formed and has reached the axis of symmetry, 
as for example in figure \ref{F:stages}(b).
Figure \ref{F:size} plots $s_\psi$ as a function of $d$,
for $p=0,1/2,1,2$, as indicated. 
After an initial transition period, 
the curves for all $p$ approach the same common curve.
For approximately $d>0.5$, this curve 
closely satisfies 
\begin{equation}
s_{\psi}\sim d^{2/3}
\end{equation}
and is in good agreement with the observations
by Taneda \& Honji \cite{taneda71} based on laboratory 
experiments.
The initial transition period is not 
a viscous or finite plate effect and thus the scaling 
does not hold asymptotically as $d\to0$ in any particular
limit. The size does reflect the growth of the vertical
vortex displacement $y_c$, which also scales as 
$d^{2/3}$. 

\section{Summary}\label{sec:conc}

This paper presents a numerical study of viscous flow past
a finite flat plate moving with accelerated velocity 
$U=t^p$ ($\Uhat=a\that^p$) in direction normal to itself. The focus is on 
the effect of the parameter $p$, for $p=0,1/2,1,2$,
for fixed value of the Reynolds number, $Re=500$.
All results are computed and presented in a reference 
frame fixed on the plate.
Most results are reported as functions of 
the plate displacement $d$, which was found to more
concisely reflect the dependence on $p$. 
We report on the vorticity contours, velocity profiles and
streaklines at fixed displacement $d$, and on the evolution of
the vortex core trajectory, maximum vorticity, and circulation 
as functions of $d$.

At fixed displacement $d$, the acceleration parameter $p$ affects the vorticity
distribution within the core. For example, at $d=1$, 
for larger values of $p$,
the outer turns of the shear layer rollup are
stronger, and the vorticity profiles near
the center are flatter.
The spiral roll-up of particle streaklines is more concentrated 
near the center, with fewer outer turns.  
The streaklines and the vortex center are in good agreement
with available experimental data\cite{pullinperry80} for $p=0.45$.



Four stages of the vorticity evolution, as proposed by Lucchini \& Tognaccini
\cite{luchini02}, can be identified.
In the initial Rayleigh stage 
the vorticity consists of an almost uniform layer around the whole plate,
including the tip, with no apparent separation.
This stage lasts for a time $T_R$ 
that scales surprisingly well with $p$, as $p^3$, and vanishes as $p\to0$. 
In the second stage a recirculation region has formed near
the plate tip, with an enclosed rotation center whose trajectory 
transitions towards self-similar growth. 
The third stage is defined by self-similar growth.
In the fourth stage the finite plate length 
significantly affects the flow and the trajectory departs from self-similar.
We note that for $p>0$ the local vorticity maximum within the starting
vortex forms much after the center of rotation, and does not
grow self-similarly but instead, oscillates in time. 

Several scaling laws are observed that follow from dimensional analysis 
arguments.
At early times, when the vortex size is small relative to the 
plate length, the flow is expected to behave as flow
past a semi-infinite plate in which the plate length is
absent. For viscous flow, dimensional analysis yields
the self-similar streamfunction given by equation (9),
from which it follows that length scales 
and vorticity values behave as 
\begin{equation}
L\sim\sqrt{\nu t}~,\quad \omega\sim t^p/(\nu t)^{3/4}~,
\label{E:viscscaling}
\end{equation}
respectively.
For inviscid flow, dimensional analysis 
implies that length scales and shed circulation grow as
\begin{equation}
L\sim t^{2(1+p)/3}~,\quad \Gamma\sim t^{(1+4p)/3}
\label{E:inviscscaling}
\end{equation}
respectively. For the viscous vortex separation computed here
some quantities closely satisfy the
viscous scaling, while others are largely independent of 
viscosity and closely satisfy the inviscid scaling. 

For example, 
the boundary layer thickness 
closely follows the viscous scaling laws until at least $d=0.1$.
The maximum core vorticity for $p=0$ is in excellent agreement
with the viscous laws as well, for practically the whole range
computed $d\in[10^{-4},1]$ (see also \cite{xunitsche14}). 
For $p>0$, the viscous scaling
is visible for a short time interval only, early on. 
The vortex center trajectory and circulation on the other
hand are largely independent of viscosity and closely follow
the inviscid laws after the initial transition time, 
until relatively large times with $d\approx 1$.
As a result, the vortex size, defined as the height of the recirculation
region on the axis, also follows the inviscid scaling
after it has reached a value comparable to the vertical vortex displacement.

The changes in the computed solutions 
between $p=0$ and $p=0.5$ are much larger than between
$p=0.5$ and $p=2$. 
The flow behaviour for $p\in(0,0.5)$ remains 
to be studied more closely,  as does the dependence
on $Re$ for $p>0$.

%

\end{document}